\documentclass[letter]{aa}
\usepackage{amsmath,graphicx,booktabs,txfonts}
\usepackage{natbib,url,hyperref,balance}
\usepackage{CJK}
\hypersetup{colorlinks=true,citecolor=blue,urlcolor=blue,linkcolor=blue}
\newcommand{\myemail}{\email{\href{mailto:yangcht@gmail.com}{yangcht@gmail.com}}}

\newcommand{\lya}{Ly$\alpha$}
\newcommand{\cii}{[C\,{\sc ii}]}
\newcommand{\ci}{[C\,{\sc i}]}
\newcommand{\nii}{[N\,{\sc ii}]}
\newcommand{\oi}{[O\,{\sc i}]}

\newcommand{\kms}{km\,s$^{-1}$}
\newcommand{\Jykms}{Jy\,\kms}
\newcommand{\Lp}{L^{\prime}}
\newcommand{\Lpunit}{K\,\kms\,pc$^{2}$}
\newcommand{\Msun}{M_{\odot}}
\newcommand{\Lsun}{L_{\odot}}
\newcommand{\Gnot}{G_{0}}
\newcommand{\aco}{\alpha_{\rm CO}}
\newcommand{\Mmol}{M_{\rm mol}}
\newcommand{\Mstar}{M_\star}
\newcommand{\tdep}{t_{\rm dep}}
\newcommand{\un}[1]{{\scriptsize\,[#1]}}

\begin{document}
\begin{CJK*}{UTF8}{gbsn}
\title{Low-excitation molecular gas in a Ly$\alpha$-associated companion of the $z=4.7$ quasar in BR1202$-$0725}
\author
{
  \href{https://orcid.org/0000-0002-8117-9991}{C. Yang(杨辰涛)}\inst{1,2}        \and
  \href{https://orcid.org/0000-0002-9931-1313}{S. Muller}\inst{3}               \and
  \href{https://orcid.org/0000-0003-1939-5885}{M. D. Lehnert}\inst{4}           \and \\
  \href{https://orcid.org/0000-0002-5268-2221}{T J L C Bakx}\inst{2}            \and
  \href{https://orcid.org/0000-0003-2027-8221}{P. Cox}\inst{5}                  \and
  \href{https://orcid.org/0000-0002-0267-9024}{G. C. Jones}\inst{6,7}           \and
  \href{https://orcid.org/0000-0002-7821-8873}{K. Knudsen}\inst{2}              \and
  \href{https://orcid.org/0000-0002-5941-5214}{C. Mazzucchelli}\inst{8}         \and
  \href{https://orcid.org/0000-0002-4005-9619}{A. Omont}\inst{5}                \and
  \href{https://orcid.org/0000-0002-7299-2876}{Zhi-Yu Zhang(张智昱)}\inst{9,10} 
}

\institute{
Independent Researcher, Gothenburg, Sweden \myemail
\and
Department of Physics and Astronomy, Chalmers University of Technology, SE-412 96 Gothenburg, Sweden
\and
Department of Physics and Astronomy, Chalmers University of Technology, Onsala Space Observatory, SE-43992 Onsala, Sweden
\and
Univ. Lyon, Univ. Lyon1, ENS de Lyon, CNRS, Centre de Recherche Astrophysique de Lyon UMR5574, 69230 Saint-Genis-Laval, France
\and
Institut d'Astrophysique de Paris, UMR7095 CNRS \& Sorbonne Universit\'e (UPMC), 98 bis bd Arago, 75014 Paris, France
\and
Kavli Institute for Cosmology, University of Cambridge, Madingley Road, Cambridge CB3 0HA, UK
\and
Cavendish Laboratory, University of Cambridge, 19 JJ Thomson Avenue, Cambridge CB3 0HE, UK
\and
Instituto de Estudios Astrof\'isicos, Facultad de Ingenier\'ia y Ciencias, Universidad Diego Portales, Avenida Ej\'ercito Libertador 441, Santiago, Chile
\and
School of Astronomy and Space Science, Nanjing University, Nanjing 210023, PR China
\and
Key Laboratory of Modern Astronomy and Astrophysics (Nanjing University), Ministry of Education, Nanjing 210023, PR China
}

\authorrunning{C. Yang et al.}
\titlerunning{Molecular gas in a \lya-associated companion of BR1202$-$0725}
\date{Received ; accepted }

\abstract{Quasar companions at high redshifts are usually gas-rich star-forming galaxies whose molecular gas masses ($\Mmol$) are estimated from \cii, dust emission or a single mid-$J$ CO line. We detect CO(5\mbox{--}4) and \ci(1\mbox{--}0) with ALMA in LAE\mbox{-}2, a \lya-associated companion of the $z=4.7$ quasar in BR1202$-$0725, and CO(5\mbox{--}4) at the location near LAE\mbox{-}1, the \lya\ emitter between the quasar and the submillimeter galaxy (SMG). With the CO(2\mbox{--}1) flux from the literature, LAE\mbox{-}2 has an $\Lp_{\rm CO(5-4)}/\Lp_{\rm CO(2-1)}$ ratio of $0.22\pm0.04$, as in high-redshift main-sequence galaxies but four times lower than in the quasar host and the SMG. A non-LTE model places the CO(5\mbox{--}4) excitation temperature only 4\,K above the cosmic microwave background (CMB), yet the same gas would show an even lower ratio against the $z=0$ CMB. The CMB-corrected $\Mmol$ from \ci, $\approx9\times10^{10}\,\Msun$, agrees with the independent \cii\ estimate, whereas CO(5\mbox{--}4) gives a four times lower mass if assumed thermalized at 40\,K with the same CO-to-H$_2$ conversion factor. Combining our 3\,mm measurement with archival far-infrared photometry, we find cold ($\approx25$\,K) dust and main-sequence star formation, with a gas-to-stellar mass ratio of $\approx4$ and a depletion time of $\approx0.7$\,Gyr. Its star formation can supply the far-ultraviolet field that the \ci, \cii, and CO lines require. Quasar companions can thus retain massive, weakly excited molecular gas reservoirs that mid-$J$ CO surveys can miss.\looseness=-1}
\keywords{galaxies: high-redshift -- galaxies: ISM -- infrared: galaxies -- submillimeter: galaxies -- radio lines: ISM -- ISM: molecules}
\maketitle

\section{Introduction}\label{sec:intro}

Luminous quasars at high redshifts often reside in dense environments: many have gas-rich, star-forming companions within a few tens of kpc \citep{Decarli2017,Wang2026}, some of which are interacting \citep{Kade2023}. These companions hold part of the cold molecular gas that fuels the build-up of stellar mass in the group \citep{Tacconi2020}, and they are exposed to the quasar's radiation. Quasar photoionization is seen in the ionized gas of a close companion \citep{Decarli2024}, but whether the quasar alters the condition of the molecular gas of its companions is open: in one system at $z\approx2.7$, quasar radiation has reduced the molecular gas seen in absorption in a companion a few kpc away to small, dense cloudlets \citep{Balashev2025}, and companions inside the virial radii of $z\approx3$ quasars have lower gas-to-stellar mass ratios than those outside \citep{Ritter2026}. Measuring both the molecular gas mass ($\Mmol$) and the excitation of those companion galaxies is difficult. The former is often estimated from \cii\ \citep[e.g.][]{Aravena2024,Bakx2024}, dust emission \citep{Scoville2016} or a single mid-$J$ CO line, which requires an excitation correction \citep[e.g.][]{Ono2022} besides the CO-to-H$_2$ conversion factor, $\aco$; the excitation ranges from sub-thermal in main-sequence disks to near-thermal in starbursts and quasar hosts \citep{Daddi2015,Boogaard2020}. At $z\approx4.7$, the 15.5\,K cosmic microwave background (CMB) also alters the excitation and the line contrast \citep{daCunha2013,Zhang2016}. Multi-$J$ CO lines constrain the excitation, and \ci\ provides an independent $\Mmol$ estimate, but such data on quasar companions are rare \citep[e.g.][]{Pensabene2021}.

The galaxy group BR1202$-$0725 at $z=4.69$ provides an opportunity for such measurements. It contains a hyperluminous quasar ($L_{\rm bol}\approx10^{48}$\,erg\,s$^{-1}$; \citealt{Yu2021,Zamora2025}) and a submillimeter galaxy (SMG), with \cii-based systemic redshifts $z_{\rm QSO}=4.6942$ and $z_{\rm SMG}=4.6915$, respectively \citep{Carilli2013}, both with measured CO ladders \citep{Omont1996,Salome2012,Wagg2014}, and two companions with associated \lya\ emission (hereafter \lya-associated; Fig.~\ref{fig:positions}). LAE\mbox{-}1 \citep{Hu1996,Fontana1996} is a metal-poor \lya\ emitter \citep[LAE;][]{Ouchi2020,Zamora2025}, and LAE\mbox{-}2 is a dusty \citep{Carilli2013,Lehnert2020}, \cii-bright \citep{Carilli2013,Lelli2026} galaxy detected in CO(2\mbox{--}1) \citep{Jones2016}, whose \lya\ emission \citep{Hu1997} has been attributed both to in situ star formation \citep{Williams2014} and to the quasar halo \citep{Drake2020}.

In this Letter, we report CO(5\mbox{--}4), \ci(1\mbox{--}0) and 3\,mm continuum observations of both companions from a deep ALMA Band~3 survey (Yang et al., in prep.). Together with published data, they yield the excitation and gas reservoir of LAE\mbox{-}2, which we compare with main-sequence galaxies, the quasar host, and the SMG within BR1202$-$0725. Appendices~\ref{app:spectra}--\ref{app:comparison} describe, in turn, the spectral extraction, the kinematics and dynamical masses, the excitation model, the dust and PDR models, and the comparison samples. We adopt a flat $\Lambda$CDM cosmology with $H_0=67.4$\,\kms\,Mpc$^{-1}$ and $\Omega_{\rm m}=0.315$ \citep{Planck2020}, so that 1\arcsec\ is 6.62\,kpc at $z=4.6942$; a \citet{Kroupa2001} initial mass function (IMF) with star formation rates ${\rm SFR_{IR}}=1.48\times10^{-10}\,L_{\rm IR}/\Lsun$ \citep{Kennicutt2012}; $\Mmol$ including helium ($\times1.36$); and $\kappa_{850}=0.077$\,m$^{2}$\,kg$^{-1}$ \citep{Dunne2000} for dust masses.

\section{Observations and measurements}\label{sec:obs}

We observed BR1202$-$0725 with the ALMA 12-m array between Nov 2019 and Jan 2020 (2019.1.00205.S, PI: C.~Yang), obtaining 12.5\,h on source in four Band~3 tunings spanning 84.0--115.9\,GHz (Fig.~\ref{fig:fullspectra}), with baselines of 15--500\,m. The data were calibrated with the ALMA pipeline in CASA 5.6.1-8. We extracted spectra by fitting, in the $uv$-plane, six fixed-position point sources (Fig.~\ref{fig:positions}) simultaneously to each visibility channel with \mbox{UVMULTIFIT} (App.~\ref{app:spectra}). Fluxes, luminosities, and line ratios are observed values unless marked as CMB-corrected; line-based $\Mmol$ values use the corrected luminosities (App.~\ref{app:radex}). Upper limits are $3\sigma$ unless otherwise specified.

\begin{figure}[t]
\;\;\;\includegraphics[width=80mm]{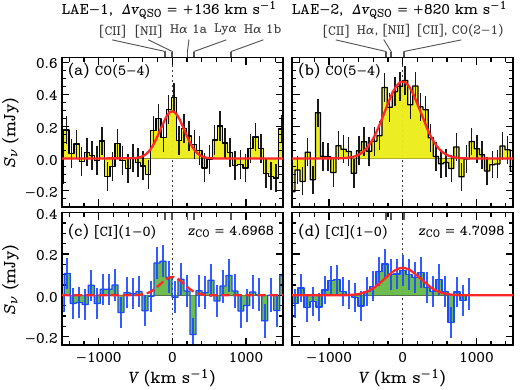}
\caption{CO(5\mbox{--}4) (top) and \ci(1\mbox{--}0) (bottom) spectra at the LAE\mbox{-}1 position (left) and of LAE\mbox{-}2 (right) in three-channel bins (70 and 81\,\kms\ for CO and \ci) vs. velocity relative to $z_{\rm CO}$ (given in the lower panels). Red: Gaussian fits (CO) and amplitude-only fits with the CO profile (\ci); dashed: undetected. Ticks: redshifts of other lines (App.~\ref{app:positions}).}\label{fig:spectra}
\end{figure}

\begin{table}[t]
\caption{Line and continuum measurements.}\label{tab:main}
\centering\footnotesize
\setlength{\tabcolsep}{3pt}
\renewcommand{\arraystretch}{0.92}
\begin{tabular}{lcc}
\toprule
Quantity & LAE\mbox{-}1 position\tablefootmark{*} & LAE\mbox{-}2\\
\midrule
$z_{\rm CO(5-4)}$ & $4.6968\pm0.0007$ & $4.7098\pm0.0005$\\
$\Delta v_{\rm QSO}$\un{\kms} & $+136\pm35$ & $+820\pm25$\\
FWHM$_{\rm CO(5-4)}$\un{\kms} & $384\pm85$ & $571\pm63$\\
$I_{\rm CO(5-4)}$\un{\Jykms} & $0.120\pm0.024$ & $0.291\pm0.030$\\
$I_{\rm [CI](1-0)}$\un{\Jykms}\tablefootmark{a} & $<0.089$ & $0.081\pm0.021$\\
$S_{100\,\rm GHz}$\un{$\mu$Jy}\tablefootmark{b} & $<31$ & $26.4\pm2.3$\\
\midrule[0.3pt]
$\Lp_{\rm CO(5-4)}$\un{$10^{9}$\,\Lpunit} & $4.0\pm0.8$ & $9.9\pm1.0$\\
$r_{52}$\tablefootmark{c} & -- & $0.22\pm0.04$\\
\bottomrule
\end{tabular}
\tablefoot{Observed values of LAE\mbox{-}1 and LAE\mbox{-}2. Derived properties: Table~\ref{tab:scenarios}. \tablefoottext{*}{Fixed extraction position (App.~\ref{app:spectra}); line likely from gas between the quasar and the SMG.} \tablefoottext{a}{Amplitude for the CO line profile; LAE\mbox{-}1 position: flux $+3\sigma$.} \tablefoottext{b}{LAE\mbox{-}1 position: flux $+3\sigma$, including the tuning scatter.} \tablefoottext{c}{With the CO(2\mbox{--}1) flux of \citet{Jones2016}, extrapolated to the full line with the CO(5\mbox{--}4) profile ($f_{\rm win}=0.894$; App.~\ref{app:radex}).}}
\end{table}

\section{Results}\label{sec:results}

\subsection{LAE\mbox{-}2: CO excitation and \ci\ emission}\label{sec:lae2}

\begin{figure}[t]
\;\includegraphics[width=85mm]{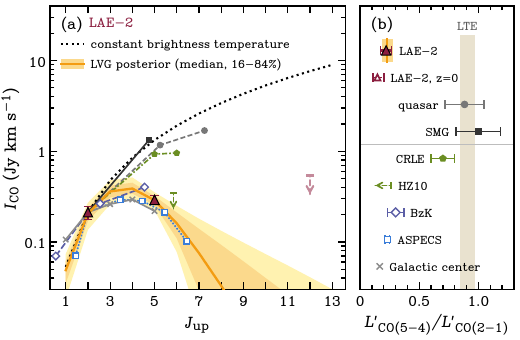}
\caption{CO excitation. (a)~LAE\mbox{-}2: CO(2\mbox{--}1) \citep{Jones2016} and CO(5\mbox{--}4) (triangles), the CO(12\mbox{--}11) limit \citep[light arrow;][]{Lee2021} and the LVG posterior (outer band: 2.5--97.5\%). Comparison ladders (symbols as in b; ASPECS: LVG model) are scaled to LAE\mbox{-}2 at CO(2\mbox{--}1) and offset by up to 0.45 in $J_{\rm up}$ for clarity. (b)~Observed $r_{52}$. Top: LAE\mbox{-}2 (box: LVG posterior, 16--84\%), its fixed-gas prediction for $z=0$, the quasar host and the SMG. Bottom: CRLE, HZ10, BzK disks, the ASPECS stacked spectra and the Galactic center (Apps.~\ref{app:radex} and \ref{app:comparison}).}\label{fig:sled}
\end{figure}

We detect CO(5\mbox{--}4) at 9.6$\sigma$ and \ci(1\mbox{--}0) at $\approx4\sigma$ in LAE\mbox{-}2, with the \ci\ centroid and width fixed to the CO profile (Fig.~\ref{fig:spectra}; Table~\ref{tab:main}). With the CO(2\mbox{--}1) flux of \citet{Jones2016}, extrapolated to the full line (App.~\ref{app:radex}), we derive $r_{52}\equiv\Lp_{\rm CO(5-4)}/\Lp_{\rm CO(2-1)}=0.22\pm0.04$ (Fig.~\ref{fig:sled}), which matches main-sequence disks at $z\approx1$--2, such as the BzK galaxies \citep{Daddi2015} and the ASPECS stacked spectra \citep{Boogaard2020}. In BR1202$-$0725, the quasar host and the SMG have $r_{52}=0.88\pm0.16$ and $1.00\pm0.19$ \citep{Salome2012,Wagg2014}, four times higher and close to the value for optically thick, thermalized CO (0.85--0.97 for 20--100\,K; the LTE band in Fig.~\ref{fig:sled}b).
To separate the state of the gas from the effect of the CMB, we fitted CO(2\mbox{--}1), CO(5\mbox{--}4) and the CO(12\mbox{--}11) limit \citep{Lee2021} with a one-component large-velocity-gradient (LVG) model (App.~\ref{app:radex}). The model requires sub-thermal CO(5\mbox{--}4), with an excitation temperature of $19.5^{+1.8}_{-1.4}$\,K, only 4\,K above the CMB. Nevertheless, the CMB cannot account for the low ratio: the same gas would show an even lower $r_{52}=0.15^{+0.06}_{-0.04}$ against the $z=0$ CMB (Fig.~\ref{fig:sled}b), so the weak CO(5\mbox{--}4) line reflects the state of the gas. The posterior favors low densities, $\log(n_{\rm H_2}/{\rm cm^{-3}})=2.75^{+0.71}_{-0.50}$, and constrains $T_{\rm kin}$ only weakly ($126^{+308}_{-90}$\,K), due to the temperature--density degeneracy.

\subsection{LAE\mbox{-}2: dust continuum}\label{sec:dust}

A new analysis of the ALMA observations published by \citet{Lehnert2020} and \citet{Lee2019} gives total fluxes of $579\pm34$\,$\mu$Jy at 223\,GHz and $1.62\pm0.45$\,mJy at 426\,GHz (App.~\ref{app:dustpdr}). Their ratio is $2.8\pm0.9$; with the higher-frequency measurement near the rest-frame dust peak, this ratio already favors cold dust. We fitted a CMB-corrected, optically thin modified blackbody to all six fluxes between 100 and 426\,GHz with flat priors on $T_{\rm d}$ (10--90\,K) and $\beta$ (1.5--2.5). We obtain $T_{\rm d}=25^{+5}_{-3}$\,K, $\beta=2.36^{+0.10}_{-0.19}$, $\log(L_{\rm IR}/\Lsun)=11.92^{+0.18}_{-0.13}$ (8--1000\,$\mu$m) and ${\rm SFR_{IR}}=122^{+64}_{-32}$\,$\Msun$\,yr$^{-1}$. The temperature is constrained jointly with $\beta$, which approaches its upper bound, but $L_{\rm IR}$ varies by less than 0.1\,dex among the other fits with free $\beta$ and flat priors (18--37\,K). Warmer dust, as in other star-forming galaxies at high redshift \citep[41--47\,K;][]{Bethermin2020,Sommovigo2022}, is not favored: fixing $T_{\rm d}=44$\,K raises $\chi^2$ from 2.4 to 12.3. The infrared SFR agrees with the SED-based SFR of \citet{Lelli2026}, $95^{+186}_{-44}$\,$\Msun$\,yr$^{-1}$, and places LAE\mbox{-}2 on the main sequence \citep[$\delta_{\rm MS}\equiv\log({\rm SFR/SFR_{MS}})=0.12$\,dex;][]{Speagle2014}.

\subsection{CO(5\mbox{--}4) at the LAE\mbox{-}1 position}\label{sec:lae1}

At the LAE\mbox{-}1 position, 1.7\arcsec\ (11.4\,kpc) from the quasar, we recover a 5.0$\sigma$ CO(5\mbox{--}4) signal at $\Delta v_{\rm QSO}=+136\pm35$\,\kms, with $I_{\rm CO(5-4)}=0.120\pm0.024$\,\Jykms. Its recovery in our short-baseline CO observations, without a compact counterpart in the higher-resolution \cii\ data of \citet{Lelli2026}, suggests diffuse molecular gas between the quasar and the SMG. Neither the quasar host nor the SMG accounts for it (3.4--4.5$\sigma$ in joint fits), and it fades toward the stellar peaks of LAE\mbox{-}1, with which it may still be associated (App.~\ref{app:spectra}). Neither \ci\ nor the dust continuum is securely detected at this position (Tables~\ref{tab:main} and \ref{tab:cont}).

\section{Discussion}\label{sec:discussion}

\subsection{LAE\mbox{-}2: molecular gas mass, depletion time and growth}\label{sec:gas}

The carbon lines give consistent gas masses, and the implied $\aco$ is a third of the Milky Way disk value \citep{Bolatto2013}. The CMB-corrected \ci\ line gives $\Mmol=9.2^{+8.8}_{-4.3}\times10^{10}\,\Msun$ with $\alpha_{\rm CI}=17\,\Msun\,$(\Lpunit)$^{-1}$ \citep{Dunne2022} and the \cii\ line gives $10.8^{+10.7}_{-5.4}\times10^{10}\,\Msun$ with the calibration of \citet{Zanella2018}, whereas the adopted dust calibration gives a mass lower by a factor of $\approx2$. With the CMB-corrected $\Lp_{\rm CO(1-0)}=6.7\times10^{10}$\,\Lpunit\ derived from our LVG model, the \ci-based mass implies $\aco=1.4\,\Msun\,$(\Lpunit)$^{-1}$, a starburst-like value despite the disk-like excitation. Varying $\aco$ from 0.8 to 4.4 gives (3--45)$\times10^{10}\,\Msun$, including the luminosity uncertainty, but the dynamical mass and the oxygen budget of LAE\mbox{-}2 argue against the upper end (Apps.~\ref{app:positions} and \ref{app:dustpdr}). If the CO were instead assumed to be thermalized at 40\,K, CO(5\mbox{--}4) would give a mass four times lower for the same $\aco$. We therefore adopt the \ci-based mass, from an optically thin line that needs no assumption on $\aco$; with $\log(M_{\rm dust}/\Msun)=8.38^{+0.19}_{-0.17}$, the gas-to-dust ratio (GDR) is $\approx380$.

\begin{figure*}[t]
\centering\includegraphics[width=150mm]{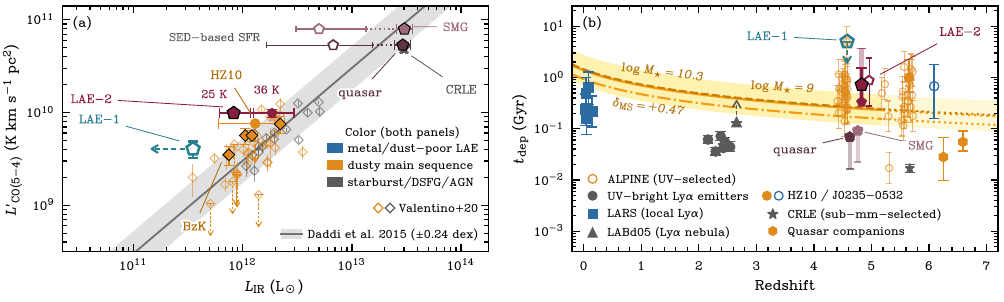}
\caption{(a) Observed $\Lp_{\rm CO(5-4)}$ vs. $L_{\rm IR}$. Pentagons: the BR1202$-$0725 group (dark bars: 16--84\%); LAE\mbox{-}2: fiducial (25\,K) and warm-prior (36\,K) fits; quasar host and SMG: filled for the $L_{\rm IR}$ of \citet{Lu2017}, open for the IR-equivalent luminosities of their SED-based SFRs. (b) $\tdep$ vs. redshift and the relation of \citet{Tacconi2020} for the fiducial (solid, $\pm0.3$\,dex) and warm-prior (dash-dotted) SFRs of LAE\mbox{-}2 and on the main sequence at $\log\Mstar=9$ (dashed), dotted beyond $z=5.2$. LAE\mbox{-}2: IR-based (filled; large: fiducial) and SED-based (open) values; pale bars: the $\aco$ range of Sect.~\ref{sec:gas} and down to $\aco=0.8$ for the quasar host and the SMG. LAE\mbox{-}1: upper limits, except $\Lp_{\rm CO(5-4)}$. Samples and references: App.~\ref{app:comparison}.}\label{fig:consumption}
\end{figure*}

LAE\mbox{-}2 is therefore gas-rich and consumes its gas slowly. Using the stellar mass reported by \citet{Lelli2026}, $\log(\Mstar/\Msun)=10.33^{+0.20}_{-0.22}$, its gas-to-stellar mass ratio is $\mu_{\rm gas}=\Mmol/\Mstar=4.4^{+5.7}_{-2.4}$, and its depletion time ($\tdep=\Mmol/{\rm SFR}$) for the cold-dust SFR is $0.72^{+0.80}_{-0.38}$\,Gyr. Both are about three times the predictions of the relation of \citet{Tacconi2020}, $\mu_{\rm gas}=1.7$ and $\tdep=0.25$\,Gyr (Figs.~\ref{fig:gasfraction} and \ref{fig:consumption}b); the offsets shrink to $\leq0.15$\,dex with the dust-based mass. Its $\tdep$ is also 8--11 times longer than in the quasar host and the SMG if star formation powers all their infrared emission; with the SED-based SFRs of \citet{Lelli2026}, their median $\tdep$ values remain shorter (0.30 and 0.52\,Gyr vs. 0.88\,Gyr for LAE\mbox{-}2 on the same basis), but the uncertainty intervals overlap (App.~\ref{app:comparison}). Consistent with this slow consumption, CO(5\mbox{--}4), although weak relative to CO(2\mbox{--}1), is bright relative to $L_{\rm IR}$: LAE\mbox{-}2 lies $\approx0.6$\,dex above the relation of \citet{Daddi2015} (scatter 0.24\,dex; Fig.~\ref{fig:consumption}a).

\subsection{LAE\mbox{-}2 in the quasar environment}\label{sec:env}

LAE\mbox{-}2 lies at a projected 14.5\,kpc and $+820$\,\kms\ from the quasar, below the escape speed of an assumed $10^{12.5}$--$10^{13}\,\Msun$ halo (App.~\ref{app:positions}), and its dust shows a small, probably tidal tail toward the quasar \citep{Lelli2026}; it may therefore be a bound satellite near pericenter. Does the quasar affect its gas?

The line ratios of LAE\mbox{-}2 do not require quasar heating, although they do not exclude it. Photon-dominated region (PDR) models \citep{Wolfire2022} fit the \cii, \ci\ and CO(5\mbox{--}4) lines of LAE\mbox{-}2 with a dense component, $n_{\rm H}\approx10^{4.5}$\,cm$^{-3}$, in a far-ultraviolet (FUV) field of $\log\Gnot\approx2.8$ (Habing units) when the cold-dust $L_{\rm IR}$ is used as a constraint; this component supplies only 26--37\% of the CO(2\mbox{--}1) luminosity. The field is weaker than both that of the galaxy's own star formation, averaged over the dust area ($\log\Gnot\approx4.1$), and the unattenuated quasar field at 14.5\,kpc ($\log\Gnot=3.65$); complete absorption of the quasar light intercepted by the compact dust would supply at most 6\% (FUV) or 30\% (bolometric) of $L_{\rm IR}$ (App.~\ref{app:dustpdr}). Whereas \citet{Balashev2025} infer highly excited cloudlets in the quasar-illuminated absorbing gas of their $z\approx2.7$ companion, the low-$J$ CO emission of LAE\mbox{-}2 is dominated by weakly excited gas.

The median $\tdep$ exceeds the orbital time at 15--30\,kpc in the assumed halo (0.15--0.45\,Gyr), so the reservoir could supply substantial fuel if LAE\mbox{-}2 merges with the quasar host before it is depleted: its present $\Mmol$ is a quarter to a third of that of the host, whereas its stellar mass is only about a thirteenth of the host's. With the conventions of \citet{Ritter2026}, LAE\mbox{-}2 has $\Mmol/\Mstar=0.20$--0.43, comparable to or above their median for companions inside the virial radii of $z\approx3$ quasars; it is thus not unusually gas-poor among such companions (App.~\ref{app:comparison}).

\section{Conclusions}\label{sec:conclusions}

We have presented ALMA observations of CO(5\mbox{--}4), \ci, and the 3\,mm continuum toward LAE\mbox{-}1 and LAE\mbox{-}2 in the $z\simeq4.7$ quasar group BR1202$-$0725. LAE\mbox{-}2 retains a substantial, weakly excited molecular gas reservoir only 14.5\,kpc in projection from the hyperluminous quasar. Its $r_{52}=0.22\pm0.04$ matches main-sequence disks and is four times lower than in the quasar host and the SMG. The \ci\ line yields $\approx9\times10^{10}\,\Msun$, and the \cii\ calibration recovers this mass, so \cii\ can be a reliable tracer of molecular gas in similar high-redshift galaxies. Combining our 3\,mm flux with the archival continuum data, we infer cold ($\approx25$\,K) dust and a main-sequence SFR of $\approx120\,\Msun$\,yr$^{-1}$. On the \ci-based mass scale, LAE\mbox{-}2 has a gas-to-stellar mass ratio of four and $\tdep\approx0.7$\,Gyr. Its star formation can supply the field required by the PDR lines, although quasar heating is not excluded. This massive reservoir coexists with the quasar without evidence for strong depletion. The CO(5\mbox{--}4) signal at the LAE\mbox{-}1 position likely arises from diffuse gas between the quasar and the SMG, tidal material or gas of LAE\mbox{-}1 offset from its stars. Changing only the assumed CO excitation to thermalized 40\,K gas would underestimate an LAE\mbox{-}2-like reservoir by a factor of four: mid-$J$ CO surveys can miss such reservoirs, whereas the carbon lines provide consistent molecular-gas mass estimates.

\begin{acknowledgements}
CY and SM acknowledge support from the Nordic ALMA Regional Centre (ARC) node based at Onsala Space Observatory. The Nordic ARC node is funded through Swedish Research Council grant No. 2019-00208. GCJ acknowledges support by the Science and Technology Facilities Council (STFC), by the ERC through Advanced Grant 695671 ``QUENCH'', and by the UKRI Frontier Research grant RISEandFALL. KK acknowledges support for the Knut and Alice Wallenberg Foundation (KAW 2023.0339). CM acknowledges support from Fondecyt Iniciacion grant 11240336 and the ANID BASAL project FB210003. Z.-Y.Z. acknowledges the support of the National Natural Science Foundation of China (NSFC) under grants No. 12533003, 1257030642. Z.-Y.Z. is supported by the Fundamental Research Funds for the Central Universities (Grant no. KG202502). This paper uses the following ALMA data: ADS/JAO.ALMA\#2019.1.00205.S, \#2019.1.01587.S, \#2015.1.01489.S and \#2013.1.00745.S. ALMA is a partnership of ESO (representing its member states), NSF (USA) and NINS (Japan), together with NRC (Canada), NSTC and ASIAA (Taiwan), and KASI (Republic of Korea), in cooperation with the Republic of Chile. The Joint ALMA Observatory is operated by ESO, AUI/NRAO and NAOJ. This work is based in part on observations made with the NASA/ESA \textit{Hubble} Space Telescope and the NASA/ESA/CSA \textit{James Webb} Space Telescope, obtained from the Mikulski Archive for Space Telescopes at the Space Telescope Science Institute, which is operated by the Association of Universities for Research in Astronomy, Inc., under NASA contracts NAS 5-26555 and NAS 5-03127; these observations are associated with programs 10417 and 3954.
\end{acknowledgements}

\let\voldthebibliography\thebibliography
\renewcommand{\thebibliography}[1]{\voldthebibliography{#1}\setlength{\itemsep}{-1pt}\setlength{\parsep}{-1pt}}
\bibliographystyle{aa_url}
\bibliography{references}

@ARTICLE{Balashev2025,
  author = {{Balashev}, Sergei and {Noterdaeme}, Pasquier and {Gupta}, Neeraj and {Krogager}, Jens-Kristian and {Combes}, Fran{\c c}oise and {L{\'o}pez}, Sebasti{\'a}n and {Petitjean}, Patrick and {Omont}, Alain and {Srianand}, Raghunathan and {Cuellar}, Rodrigo},
  title = {Quasar radiation transforms the gas in a merging companion galaxy},
  journal = {\nat},
  year = {2025},
  volume = {641},
  pages = {1137--1141},
  doi = {10.1038/s41586-025-08966-4},
  archivePrefix = {arXiv},
  eprint = {2505.15766},
  adsurl = {https://ui.adsabs.harvard.edu/abs/2025Natur.641.1137B}
}

@ARTICLE{Bethermin2020,
  author = {{B{\'e}thermin}, M. and {Fudamoto}, Y. and {Ginolfi}, M. and {Loiacono}, F. and {Khusanova}, Y. and {Capak}, P. L. and {Cassata}, P. and {Faisst}, A. and {Le F{\`e}vre}, O. and {Schaerer}, D. and {Silverman}, J. D. and {Yan}, L. and {Amorin}, R. and {Bardelli}, S. and {Boquien}, M. and {Cimatti}, A. and {Davidzon}, I. and {Dessauges-Zavadsky}, M. and {Fujimoto}, S. and {Gruppioni}, C. and {Hathi}, N. P. and {Ibar}, E. and {Jones}, G. C. and {Koekemoer}, A. M. and {Lagache}, G. and {Lemaux}, B. C. and {Moreau}, C. and {Oesch}, P. A. and {Pozzi}, F. and {Riechers}, D. A. and {Talia}, M. and {Toft}, S. and {Vallini}, L. and {Vergani}, D. and {Zamorani}, G. and {Zucca}, E.},
  title = {{The ALPINE-ALMA [CII] survey: Data processing, catalogs, and statistical source properties}},
  journal = {\aap},
  year = {2020},
  volume = {643},
  eid = {A2},
  pages = {A2},
  doi = {10.1051/0004-6361/202037649},
  adsurl = {https://ui.adsabs.harvard.edu/abs/2020A&A...643A...2B}
}

@ARTICLE{Bolatto2013,
  author = {{Bolatto}, A.~D. and {Wolfire}, M. and {Leroy}, A.~K.},
  title = {{The CO-to-H$_{2}$ Conversion Factor}},
  journal = {\araa},
  year = {2013},
  volume = {51},
  pages = {207--268},
  doi = {10.1146/annurev-astro-082812-140944},
  archivePrefix = {arXiv},
  eprint = {1301.3498},
  adsurl = {https://ui.adsabs.harvard.edu/abs/2013ARA&A..51..207B}
}

@ARTICLE{Boogaard2020,
  author = {Boogaard, Leindert A. and {van der Werf}, Paul and Weiss, Axel and Popping, Gergo and Decarli, Roberto and Walter, Fabian and others},
  title = {The {ALMA} Spectroscopic Survey in the {HUDF}: {CO} Excitation and Atomic Carbon in Star-Forming Galaxies at {$z=1-3$}},
  journal = {\apj},
  year = {2020},
  volume = {902},
  eid = {109},
  pages = {109},
  doi = {10.3847/1538-4357/abb82f},
  archivePrefix = {arXiv},
  eprint = {2009.04348},
  adsurl = {https://ui.adsabs.harvard.edu/abs/2020ApJ...902..109B}
}

@ARTICLE{Carilli2013,
  author = {{Carilli}, C.~L. and {Riechers}, D. and {Walter}, F. and {Maiolino}, R. and 
	{Wagg}, J. and {Lentati}, L. and {McMahon}, R. and {Wolfe}, A.},
  title = {{The Anatomy of an Extreme Starburst within 1.3 Gyr of the Big Bang Revealed by ALMA}},
  journal = {\apj},
  year = {2013},
  volume = {763},
  eid = {120},
  pages = {120},
  doi = {10.1088/0004-637X/763/2/120},
  archivePrefix = {arXiv},
  eprint = {1211.6973},
  adsurl = {https://ui.adsabs.harvard.edu/abs/2013ApJ...763..120C}
}

@ARTICLE{Croxall2017,
  author = {{Croxall}, K. V. and {Smith}, J. D. and {Pellegrini}, E. and {Groves}, B. and {Bolatto}, A. and {Herrera-Camus}, R. and {Sandstrom}, K. M. and {Draine}, B. and {Wolfire}, M. G. and {Armus}, L. and {Boquien}, M. and {Brandl}, B. and {Dale}, D. and {Galametz}, M. and {Hunt}, L. and {Kennicutt}, R. and {Kreckel}, K. and {Rigopoulou}, D. and {Werf}, P. van der and {Wilson}, C.},
  title = {{The Origins of [C ii] Emission in Local Star-forming Galaxies}},
  journal = {\apj},
  year = {2017},
  volume = {845},
  eid = {96},
  pages = {96},
  doi = {10.3847/1538-4357/aa8035},
  adsurl = {https://ui.adsabs.harvard.edu/abs/2017ApJ...845...96C}
}

@ARTICLE{daCunha2013,
  author = {{da Cunha}, E. and {Groves}, B. and {Walter}, F. and {Decarli}, R. and {Wei{\ss}}, A. and {Bertoldi}, F. and {Carilli}, C. and {Daddi}, E. and {Elbaz}, D. and {Ivison}, R. and {Maiolino}, R. and {Riechers}, D. and {Rix}, H.-W. and {Sargent}, M. and {Smail}, I.},
  title = {{On the Effect of the Cosmic Microwave Background in High-redshift (Sub-)millimeter Observations}},
  journal = {\apj},
  year = {2013},
  volume = {766},
  eid = {13},
  pages = {13},
  doi = {10.1088/0004-637X/766/1/13},
  archivePrefix = {arXiv},
  eprint = {1302.0844},
  adsurl = {https://ui.adsabs.harvard.edu/abs/2013ApJ...766...13D}
}

@ARTICLE{Daddi2015,
  author = {{Daddi}, E. and {Dannerbauer}, H. and {Liu}, D. and {Aravena}, M. and {Bournaud}, F. and {Walter}, F. and {Riechers}, D. and {Magdis}, G. and {Sargent}, M. and {B{\'e}thermin}, M. and {Carilli}, C. and {Cibinel}, A. and {Dickinson}, M. and {Elbaz}, D. and {Gao}, Y. and {Gobat}, R. and {Hodge}, J. and {Krips}, M.},
  title = {{CO excitation of normal star-forming galaxies out to z = 1.5 as regulated by the properties of their interstellar medium}},
  journal = {\aap},
  year = {2015},
  volume = {577},
  eid = {A46},
  pages = {A46},
  doi = {10.1051/0004-6361/201425043},
  archivePrefix = {arXiv},
  eprint = {1409.8158},
  adsurl = {https://ui.adsabs.harvard.edu/abs/2015A&A...577A..46D}
}

@ARTICLE{Decarli2014,
  author = {{Decarli}, R. and {Walter}, F. and {Carilli}, C. and {Bertoldi}, F. and {Cox}, P. and {Ferkinhoff}, C. and {Groves}, B. and {Maiolino}, R. and {Neri}, R. and {Riechers}, D. and {Weiss}, A.},
  title = {Varying {[C II]/[N II]} Line Ratios in the Interacting System {BR1202-0725} at {$z=4.7$}},
  journal = {\apjl},
  year = {2014},
  volume = {782},
  eid = {L17},
  pages = {L17},
  doi = {10.1088/2041-8205/782/2/l17},
  archivePrefix = {arXiv},
  eprint = {1401.5076},
  adsurl = {https://ui.adsabs.harvard.edu/abs/2014ApJ...782L..17D}
}

@ARTICLE{Dessauges2020,
  author = {{Dessauges-Zavadsky}, M. and {Ginolfi}, M. and {Pozzi}, F. and {B{\'e}thermin}, M. and {Le F{\`e}vre}, O. and {Fujimoto}, S. and {Silverman}, J. D. and {Jones}, G. C. and {Vallini}, L. and {Schaerer}, D. and {Faisst}, A. L. and {Khusanova}, Y. and {Fudamoto}, Y. and {Cassata}, P. and {Loiacono}, F. and {Capak}, P. L. and {Yan}, L. and {Amorin}, R. and {Bardelli}, S. and {Boquien}, M. and {Cimatti}, A. and {Gruppioni}, C. and {Hathi}, N. P. and {Ibar}, E. and {Koekemoer}, A. M. and {Lemaux}, B. C. and {Narayanan}, D. and {Oesch}, P. A. and {Rodighiero}, G. and {Romano}, M. and {Talia}, M. and {Toft}, S. and {Vergani}, D. and {Zamorani}, G. and {Zucca}, E.},
  title = {{The ALPINE-ALMA [C II] survey}},
  journal = {\aap},
  year = {2020},
  volume = {643},
  eid = {A5},
  pages = {A5},
  doi = {10.1051/0004-6361/202038231},
  adsurl = {https://ui.adsabs.harvard.edu/abs/2020A&A...643A...5D}
}

@ARTICLE{Dessauges2025,
  author = {Dessauges-Zavadsky, M. and Marques-Chaves, R. and Schaerer, D. and Xiao, M.-Y. and Colina, L. and {\'{A}lvarez-M\'{a}rquez}, J. and {P\'{e}rez-Fournon}, I.},
  title = {Unveiling Dust, Molecular Gas, and High Star-formation Efficiency in Extremely {UV} Bright Star-forming Galaxies at {$z\sim2.1-3.6$}},
  journal = {\aap},
  year = {2025},
  volume = {693},
  eid = {A17},
  pages = {A17},
  doi = {10.1051/0004-6361/202451832},
  archivePrefix = {arXiv},
  eprint = {2410.11121},
  adsurl = {https://ui.adsabs.harvard.edu/abs/2025A&A...693A..17D}
}

@ARTICLE{Drake2020,
  author = {Drake, A. B. and Walter, F. and Novak, M. and Farina, E. P. and Neeleman, M. and Riechers, D. and Carilli, C. and Decarli, R. and Mazzucchelli, C. and Onoue, M.},
  title = {The Ionized- and Cool-gas Content of the {BR1202-0725} System as Seen by {MUSE} and {ALMA}},
  journal = {\apj},
  year = {2020},
  volume = {902},
  eid = {37},
  pages = {37},
  doi = {10.3847/1538-4357/aba832},
  archivePrefix = {arXiv},
  eprint = {2007.14221},
  adsurl = {https://ui.adsabs.harvard.edu/abs/2020ApJ...902...37D}
}

@ARTICLE{Dunne2000,
  author = {{Dunne}, L. and {Eales}, S. and {Edmunds}, M. and {Ivison}, R. and {Alexander}, P. and {Clements}, D. L.},
  title = {The {SCUBA} Local Universe Galaxy Survey -- {I}. First measurements of the submillimetre luminosity and dust mass functions},
  journal = {\mnras},
  year = {2000},
  volume = {315},
  pages = {115--139},
  doi = {10.1046/j.1365-8711.2000.03386.x},
  archivePrefix = {arXiv},
  eprint = {astro-ph/0002234},
  adsurl = {https://ui.adsabs.harvard.edu/abs/2000MNRAS.315..115D}
}

@ARTICLE{Dunne2022,
  author = {{Dunne}, L and {Maddox}, S J and {Papadopoulos}, P P and {Ivison}, R J and {Gomez}, H L},
  title = {Dust, {CO}, and {[C I]}: Cross-calibration of Molecular Gas Mass Tracers in Metal-rich Galaxies across Cosmic Time},
  journal = {\mnras},
  year = {2022},
  volume = {517},
  pages = {962--999},
  doi = {10.1093/mnras/stac2098},
  archivePrefix = {arXiv},
  eprint = {2208.01622},
  adsurl = {https://ui.adsabs.harvard.edu/abs/2022MNRAS.517..962D}
}

@ARTICLE{ForemanMackey2013,
  author = {{Foreman-Mackey}, Daniel and {Hogg}, David W. and {Lang}, Dustin and {Goodman}, Jonathan},
  title = {{emcee: The MCMC Hammer}},
  journal = {\pasp},
  year = {2013},
  volume = {125},
  pages = {306--312},
  doi = {10.1086/670067},
  adsurl = {https://ui.adsabs.harvard.edu/abs/2013PASP..125..306F}
}

@ARTICLE{Hu1996,
  author = {{Hu}, Esther M. and {McMahon}, Richard G. and {Egami}, Eiichi},
  title = {{Detection of a Ly$\alpha$ Emission-Line Companion to the z = 4.69 Quasar BR 1202$-$0725}},
  journal = {\apjl},
  year = {1996},
  volume = {459},
  eid = {L53},
  pages = {L53},
  doi = {10.1086/309958},
  adsurl = {https://ui.adsabs.harvard.edu/abs/1996ApJ...459L..53H}
}

@ARTICLE{Jones2016,
  author = {{Jones}, G.~C. and {Carilli}, C.~L. and {Momjian}, E. and {Wagg}, J. and 
	{Riechers}, D.~A. and {Walter}, F. and {Decarli}, R. and {Ota}, K. and 
	{McMahon}, R.},
  title = {{New Constraints on the Molecular Gas in the Prototypical HyLIRGs BRI1202-0725 and BRI1335-0417}},
  journal = {\apj},
  year = {2016},
  volume = {830},
  eid = {63},
  pages = {63},
  doi = {10.3847/0004-637X/830/2/63},
  archivePrefix = {arXiv},
  eprint = {1607.06755},
  adsurl = {https://ui.adsabs.harvard.edu/abs/2016ApJ...830...63J}
}

@ARTICLE{Kennicutt2012,
  author = {{Kennicutt}, Robert C. and {Evans}, Neal J.},
  title = {{Star Formation in the Milky Way and Nearby Galaxies}},
  journal = {\araa},
  year = {2012},
  volume = {50},
  pages = {531--608},
  doi = {10.1146/annurev-astro-081811-125610},
  adsurl = {https://ui.adsabs.harvard.edu/abs/2012ARA&A..50..531K}
}

@ARTICLE{Lee2021,
  author = {{Lee}, Minju M. and {Nagao}, Tohru and {De Breuck}, Carlos and {Carniani}, Stefano and {Cresci}, Giovanni and {Hatsukade}, Bunyo and {Kawabe}, Ryohei and {Kohno}, Kotaro and {Maiolino}, Roberto and {Mannucci}, Fillipo and {Marconi}, Alessandro and {Nakanishi}, Kouichiro and {Troncoso}, Paulina and {Umehata}, Hideki},
  title = {Dense and Warm Neutral Gas in {BR1202-0725} at {$z=4.7$} as Traced by the {[O I]} 145 Micron Line},
  journal = {\apj},
  year = {2021},
  volume = {913},
  eid = {41},
  pages = {41},
  doi = {10.3847/1538-4357/abe7ea},
  adsurl = {https://ui.adsabs.harvard.edu/abs/2021ApJ...913...41L}
}

@ARTICLE{Lelli2026,
  author = {{Lelli}, Federico and {Lin}, Lingrui and {Zhang}, Zhi-Yu and {Bacchini}, Cecilia and {Bevacqua}, Davide and {Bhangal}, Joe and {Castellano}, Marco and {De Breuck}, Carlos and {Gandolfi}, Giovanni and {Man}, Allison W. S. and {Marasco}, Antonino and {Paris}, Diego and {P{\'e}rez-D{\'i}az}, Borja and {Santini}, Paola},
  title = {{TRICEPS. I. Tracing cold gas, dust, and stars at sub-kpc scales in massive star-forming galaxies at z = 4-5}},
  journal = {arXiv e-prints},
  year = {2026},
  eid = {arXiv:2609.30375},
  pages = {arXiv:2609.30375},
  archivePrefix = {arXiv},
  eprint = {2609.30375},
  adsurl = {https://ui.adsabs.harvard.edu/abs/2026arXiv260930375L}
}

@ARTICLE{Lu2017,
  author = {{Lu}, Nanyao and {Zhao}, Yinghe and {D{\'{i}}az-Santos}, Tanio and {Xu}, C. Kevin and {Charmandaris}, Vassilis and {Gao}, Yu and {van der Werf}, Paul P. and {Privon}, George C. and {Inami}, Hanae and {Rigopoulou}, Dimitra and {Sanders}, David B. and {Zhu}, Lei},
  title = {{ALMA} {[NII]} 205 {$\mu$m} Imaging Spectroscopy of the Interacting Galaxy System {BRI 1202-0725} at Redshift 4.7},
  journal = {\apjl},
  year = {2017},
  volume = {842},
  eid = {L16},
  pages = {L16},
  doi = {10.3847/2041-8213/aa77fc},
  adsurl = {https://ui.adsabs.harvard.edu/abs/2017ApJ...842L..16L}
}

@ARTICLE{Lusso2015,
  author = {{Lusso}, E. and {Worseck}, G. and {Hennawi}, J. F. and {Prochaska}, J. X. and {Vignali}, C. and {Stern}, J. and {O'Meara}, J. M.},
  title = {{The first ultraviolet quasar-stacked spectrum at z $\simeq$ 2.4 from WFC3}},
  journal = {\mnras},
  year = {2015},
  volume = {449},
  pages = {4204--4220},
  doi = {10.1093/mnras/stv516},
  adsurl = {https://ui.adsabs.harvard.edu/abs/2015MNRAS.449.4204L}
}

@ARTICLE{MartiVidal2014,
  author = {Mart{\'i}-Vidal, I. and Vlemmings, W. H. T. and Muller, S. and Casey, S.},
  title = {{UVMULTIFIT}: A versatile tool for fitting astronomical radio interferometric data},
  journal = {\aap},
  year = {2014},
  volume = {563},
  eid = {A136},
  pages = {A136},
  doi = {10.1051/0004-6361/201322633},
  archivePrefix = {arXiv},
  eprint = {1401.4984},
  adsurl = {https://ui.adsabs.harvard.edu/abs/2014A&A...563A.136M}
}

@ARTICLE{Ohyama2004,
  author = {{Ohyama}, Youichi and {Taniguchi}, Yoshiaki and {Shioya}, Yasuhiro},
  title = {{Subaru Deep Spectroscopy of a Star-forming Companion Galaxy of BR 1202-0725 at z = 4.7}},
  journal = {\aj},
  year = {2004},
  volume = {128},
  pages = {2704--2711},
  doi = {10.1086/425887},
  adsurl = {https://ui.adsabs.harvard.edu/abs/2004AJ....128.2704O}
}

@ARTICLE{Ono2022,
  author = {{Ono}, Yoshiaki and {Fujimoto}, Seiji and {Harikane}, Yuichi and {Ouchi}, Masami and {Vallini}, Livia and {Ferrara}, Andrea and {Shibuya}, Takatoshi and {Pallottini}, Andrea and {Inoue}, Akio K. and {Imanishi}, Masatoshi and {Shimasaku}, Kazuhiro and {Hashimoto}, Takuya and {Lee}, Chien-Hsiu and {Sugahara}, Yuma and {Tamura}, Yoichi and {Kohno}, Kotaro and {Schramm}, Malte},
  title = {{ALMA Observations of CO Emission from Luminous Lyman-break Galaxies at z = 6.0293--6.2037}},
  journal = {\apj},
  year = {2022},
  volume = {941},
  eid = {74},
  pages = {74},
  doi = {10.3847/1538-4357/ac9ea6},
  archivePrefix = {arXiv},
  eprint = {2211.00793},
  primaryClass = {astro-ph.GA},
  adsurl = {https://ui.adsabs.harvard.edu/abs/2022ApJ...941...74O}
}

@ARTICLE{Ouchi2020,
  author = {{Ouchi}, Masami and {Ono}, Yoshiaki and {Shibuya}, Takatoshi},
  title = {Observations of the {Lyman-$\alpha$} Universe},
  journal = {\araa},
  year = {2020},
  volume = {58},
  pages = {617--659},
  doi = {10.1146/annurev-astro-032620-021859},
  archivePrefix = {arXiv},
  eprint = {2012.07960},
  adsurl = {https://ui.adsabs.harvard.edu/abs/2020ARA&A..58..617O}
}

@ARTICLE{Pavesi2016,
  author = {{Pavesi}, Riccardo and {Riechers}, Dominik A. and {Capak}, Peter L. and {Carilli}, Christopher L. and {Sharon}, Chelsea E. and {Stacey}, Gordon J. and {Karim}, Alexander and {Scoville}, Nicholas Z. and {Smol{\v{c}}i{\'{c}}}, Vernesa},
  title = {{ALMA REVEALS WEAK [N ii] EMISSION IN ``TYPICAL'' GALAXIES AND INTENSE STARBURSTS AT z = 5--6}},
  journal = {\apj},
  year = {2016},
  volume = {832},
  eid = {151},
  pages = {151},
  doi = {10.3847/0004-637X/832/2/151},
  archivePrefix = {arXiv},
  eprint = {1607.02520},
  primaryClass = {astro-ph.GA},
  adsurl = {https://ui.adsabs.harvard.edu/abs/2016ApJ...832..151P}
}

@ARTICLE{Pavesi2019,
  author = {{Pavesi}, Riccardo and {Riechers}, Dominik A. and {Faisst}, Andreas L. and {Stacey}, Gordon J. and {Capak}, Peter L.},
  title = {{Low Star Formation Efficiency in Typical Galaxies at z = 5--6}},
  journal = {\apj},
  year = {2019},
  volume = {882},
  eid = {168},
  pages = {168},
  doi = {10.3847/1538-4357/ab3a46},
  archivePrefix = {arXiv},
  eprint = {1812.00006},
  primaryClass = {astro-ph.GA},
  adsurl = {https://ui.adsabs.harvard.edu/abs/2019ApJ...882..168P}
}

@ARTICLE{Pensabene2021,
  author = {{Pensabene}, A. and {Decarli}, R. and {Ba{\~n}ados}, E. and {Venemans}, B. and {Walter}, F. and {Bertoldi}, F. and {Fan}, X. and {Farina}, E. P. and {Li}, J. and {Mazzucchelli}, C. and {Novak}, M. and {Riechers}, D. and {Rix}, H.-W. and {Strauss}, M. A. and {Wang}, R. and {Wei{\ss}}, A. and {Yang}, J. and {Yang}, Y.},
  title = {{ALMA multiline survey of the ISM in two quasar host--companion galaxy pairs at z > 6}},
  journal = {\aap},
  year = {2021},
  volume = {652},
  eid = {A66},
  pages = {A66},
  doi = {10.1051/0004-6361/202039696},
  adsurl = {https://ui.adsabs.harvard.edu/abs/2021A&A...652A..66P}
}

@ARTICLE{Planck2020,
  author = {{Planck Collaboration} and {Aghanim}, N. and {Akrami}, Y. and {Ashdown}, M. and {Aumont}, J. and {Baccigalupi}, C. and {Ballardini}, M. and {Banday}, A. J. and {Barreiro}, R. B. and {Bartolo}, N. and others},
  title = {{Planck 2018 results. VI. Cosmological parameters}},
  journal = {\aap},
  year = {2020},
  volume = {641},
  eid = {A6},
  pages = {A6},
  doi = {10.1051/0004-6361/201833910},
  adsurl = {https://ui.adsabs.harvard.edu/abs/2020A&A...641A...6P}
}

@ARTICLE{Puschnig2020,
  author = {{Puschnig}, J. and {Hayes}, M. and {{\"{O}}stlin}, G. and {Cannon}, J. and {Smirnova-Pinchukova}, I. and {Husemann}, B. and {Kunth}, D. and {Bridge}, J. and {Herenz}, E. C. and {Messa}, M. and {Oteo}, I.},
  title = {The {Lyman Alpha Reference Sample XI}: Efficient Turbulence Driven {Ly$\alpha$} Escape and the Analysis of {IR, CO and [C II]}},
  journal = {\aap},
  year = {2020},
  volume = {644},
  eid = {A10},
  pages = {A10},
  doi = {10.1051/0004-6361/201936768},
  archivePrefix = {arXiv},
  eprint = {2004.09142},
  adsurl = {https://ui.adsabs.harvard.edu/abs/2020A&A...644A..10P}
}

@ARTICLE{RemyRuyer2014,
  author = {{R{\'{e}}my-Ruyer}, A. and {Madden}, S. C. and {Galliano}, F. and {Galametz}, M. and {Takeuchi}, T. T. and {Asano}, R. S. and {Zhukovska}, S. and {Lebouteiller}, V. and {Cormier}, D. and {Jones}, A. and {Bocchio}, M. and {Baes}, M. and {Bendo}, G. J. and {Boquien}, M. and {Boselli}, A. and {DeLooze}, I. and {Doublier-Pritchard}, V. and {Hughes}, T. and {Karczewski}, O. {\L}. and {Spinoglio}, L.},
  title = {Gas-to-dust mass ratios in local galaxies over a 2 dex metallicity range},
  journal = {\aap},
  year = {2014},
  volume = {563},
  eid = {A31},
  pages = {A31},
  doi = {10.1051/0004-6361/201322803},
  adsurl = {https://ui.adsabs.harvard.edu/abs/2014A&A...563A..31R}
}

@ARTICLE{Ritter2026,
  author = {{Ritter}, Jelena and {Arrigoni Battaia}, Fabrizio and {Peng}, Bo and {Gonz{\'a}lez Lobos}, Jay and {Chen}, Chian-Chou and {Obreja}, Aura and {Mu{\~n}oz-Elgueta}, Nahir and {Circosta}, Chiara},
  title = {{ALMA} visits the {QSO MUSEUM}: Connecting molecular gas and the cool circumgalactic medium around 37 $z\sim3$ quasars},
  journal = {arXiv e-prints},
  year = {2026},
  eid = {arXiv:2606.30742},
  pages = {arXiv:2606.30742},
  archivePrefix = {arXiv},
  eprint = {2606.30742},
  adsurl = {https://ui.adsabs.harvard.edu/abs/2026arXiv260630742R}
}

@ARTICLE{Salome2012,
  author = {{Salom{\'e}}, P. and {Gu{\'e}lin}, M. and {Downes}, D. and {Cox}, P. and 
	{Guilloteau}, S. and {Omont}, A. and {Gavazzi}, R. and {Neri}, R.},
  title = {{BR1202-0725: an extreme multiple merger at z = 4.7}},
  journal = {\aap},
  year = {2012},
  volume = {545},
  eid = {A57},
  pages = {A57},
  doi = {10.1051/0004-6361/201219955},
  archivePrefix = {arXiv},
  eprint = {1207.6722},
  primaryClass = {astro-ph.CO},
  adsurl = {https://ui.adsabs.harvard.edu/abs/2012A&A...545A..57S}
}

@ARTICLE{Sommovigo2022,
  author = {{Sommovigo}, L and {Ferrara}, A and {Pallottini}, A and {Dayal}, P and {Bouwens}, R J and {Smit}, R and {da Cunha}, E and {De Looze}, I and {Bowler}, R A A and {Hodge}, J and {Inami}, H and {Oesch}, P and {Endsley}, R and {Gonzalez}, V and {Schouws}, S and {Stark}, D and {Stefanon}, M and {Aravena}, M and {Graziani}, L and {Riechers}, D and {Schneider}, R and {van der Werf}, P and {Algera}, H and {Barrufet}, L and {Fudamoto}, Y and {Hygate}, A P S and {Labb{\'e}}, I and {Li}, Y and {Nanayakkara}, T and {Topping}, M},
  title = {{The ALMA REBELS Survey: cosmic dust temperature evolution out to z $\sim$ 7}},
  journal = {\mnras},
  year = {2022},
  volume = {513},
  pages = {3122--3135},
  doi = {10.1093/mnras/stac302},
  adsurl = {https://ui.adsabs.harvard.edu/abs/2022MNRAS.513.3122S}
}

@ARTICLE{Tacconi2020,
  author = {{Tacconi}, Linda J. and {Genzel}, Reinhard and {Sternberg}, Amiel},
  title = {{The Evolution of the Star-Forming Interstellar Medium Across Cosmic Time}},
  journal = {\araa},
  year = {2020},
  volume = {58},
  pages = {157--203},
  doi = {10.1146/annurev-astro-082812-141034},
  adsurl = {https://ui.adsabs.harvard.edu/abs/2020ARA&A..58..157T}
}

@ARTICLE{vanderTak2007,
  author = {{van der Tak}, F. F. S. and {Black}, J. H. and {Sch{\"o}ier}, F. L. and {Jansen}, D. J. and {van Dishoeck}, E. F.},
  title = {{A computer program for fast non-LTE analysis of interstellar line spectra}},
  journal = {\aap},
  year = {2007},
  volume = {468},
  pages = {627--635},
  doi = {10.1051/0004-6361:20066820},
  adsurl = {https://ui.adsabs.harvard.edu/abs/2007A&A...468..627V}
}

@ARTICLE{Vieira2022,
  author = {{Vieira}, Daniel and {Riechers}, Dominik A. and {Pavesi}, Riccardo and {Faisst}, Andreas L. and {Schinnerer}, Eva and {Scoville}, Nicholas Z. and {Stacey}, Gordon J.},
  title = {Molecular Gas Excitation of the Massive Dusty Starburst {CRLE} and the Main-sequence Galaxy {HZ10} at z = 5.7 in the {COSMOS} Field},
  journal = {\apj},
  year = {2022},
  volume = {925},
  eid = {174},
  pages = {174},
  doi = {10.3847/1538-4357/ac403a},
  archivePrefix = {arXiv},
  eprint = {2112.11705},
  adsurl = {https://ui.adsabs.harvard.edu/abs/2022ApJ...925..174V}
}

@ARTICLE{Wagg2014,
  author = {{Wagg}, J. and {Carilli}, C. L. and {Aravena}, M. and {Cox}, P. and {Lentati}, L. and {Maiolino}, R. and {McMahon}, R. G. and {Riechers}, D. and {Walter}, F. and {Andreani}, P. and {Hills}, R. and {Wolfe}, A.},
  title = {Karl G. Jansky Very Large Array observations of cold dust and molecular gas in starbursting quasar host galaxies at $z\sim4.5$},
  journal = {\apj},
  year = {2014},
  volume = {783},
  eid = {71},
  pages = {71},
  doi = {10.1088/0004-637X/783/2/71},
  archivePrefix = {arXiv},
  eprint = {1401.1213},
  adsurl = {https://ui.adsabs.harvard.edu/abs/2014ApJ...783...71W}
}

@ARTICLE{Williams2014,
  author = {{Williams}, R. J. and {Wagg}, J. and {Maiolino}, R. and {Foster}, C. and {Aravena}, M. and {Wiklind}, T. and {Carilli}, C. L. and {McMahon}, R. G. and {Riechers}, D. and {Walter}, F.},
  title = {Constraining the Nature of Two {Ly$\alpha$} Emitters Detected by {ALMA} at {$z=4.7$}},
  journal = {\mnras},
  year = {2014},
  volume = {439},
  pages = {2096--2101},
  doi = {10.1093/mnras/stu119},
  adsurl = {https://ui.adsabs.harvard.edu/abs/2014MNRAS.439.2096W}
}

@ARTICLE{Wolfire2022,
  author = {{Wolfire}, Mark G. and {Vallini}, Livia and {Chevance}, M{\'e}lanie},
  title = {{Photodissociation and X-Ray-Dominated Regions}},
  journal = {\araa},
  year = {2022},
  volume = {60},
  pages = {247--318},
  doi = {10.1146/annurev-astro-052920-010254},
  adsurl = {https://ui.adsabs.harvard.edu/abs/2022ARA&A..60..247W}
}

@ARTICLE{Yang2014,
  author = {Yang, Yujin and Walter, Fabian and Decarli, Roberto and Bertoldi, Frank and Weiss, Axel and Dey, Arjun and Prescott, Moire K. M. and Badescu, Toma},
  title = {Pinpointing the Molecular Gas within a {Ly$\alpha$} Blob at {$z\sim2.7$}},
  journal = {\apj},
  year = {2014},
  volume = {784},
  eid = {171},
  pages = {171},
  doi = {10.1088/0004-637X/784/2/171},
  archivePrefix = {arXiv},
  eprint = {1402.2686},
  adsurl = {https://ui.adsabs.harvard.edu/abs/2014ApJ...784..171Y}
}

@ARTICLE{Yang2017,
  author = {{Yang}, C. and {Omont}, A. and {Beelen}, A. and {Gao}, Y. and {van der Werf}, P. and {Gavazzi}, R. and {Zhang}, Z.-Y. and {Ivison}, R. and {Lehnert}, M. and {Liu}, D. and {Oteo}, I. and {Gonz{\'a}lez-Alfonso}, E. and {Dannerbauer}, H. and {Cox}, P. and {Krips}, M. and {Neri}, R. and {Riechers}, D. and {Baker}, A. J. and {Micha{\l}owski}, M. J. and {Cooray}, A. and {Smail}, I.},
  title = {{Molecular gas in the Herschel-selected strongly lensed submillimeter galaxies at z $\sim$ 2--4 as probed by multi-J CO lines}},
  journal = {\aap},
  year = {2017},
  volume = {608},
  eid = {A144},
  pages = {A144},
  doi = {10.1051/0004-6361/201731391},
  adsurl = {https://ui.adsabs.harvard.edu/abs/2017A&A...608A.144Y}
}

@ARTICLE{Zamora2025,
  author = {{Zamora}, Sandra and {Venturi}, Giacomo and {Carniani}, Stefano and {Bertola}, Elena and {Parlanti}, Eleonora and {Perna}, Michele and {Arribas}, Santiago and {B{\"{o}}ker}, Torsten and {Bunker}, Andrew J. and {Charlot}, St{\'{e}}phane and {D'Eugenio}, Francesco and {Maiolino}, Roberto and {Del Pino}, Bruno Rodr{\'{i}}guez and {{\"{U}}bler}, Hannah and {Cresci}, Giovanni and {Jones}, Gareth C. and {Lamperti}, Isabella},
  title = {{GA-NIFS}: The Highly Overdense System {BR1202-0725} at {$z\sim4.7$}. A Double Active Galactic Nucleus with Fast Outflows plus Eight Companion Galaxies},
  journal = {\aap},
  year = {2025},
  volume = {702},
  eid = {A102},
  pages = {A102},
  doi = {10.1051/0004-6361/202453236},
  archivePrefix = {arXiv},
  eprint = {2412.02751},
  adsurl = {https://ui.adsabs.harvard.edu/abs/2025A&A...702A.102Z}
}

@ARTICLE{Zanella2018,
  author = {{Zanella}, A and {Daddi}, E and {Magdis}, G and {Diaz Santos}, T and {Cormier}, D and {Liu}, D and {Cibinel}, A and {Gobat}, R and {Dickinson}, M and {Sargent}, M and {Popping}, G and {Madden}, S C and {Bethermin}, M and {Hughes}, T M and {Valentino}, F and {Rujopakarn}, W and {Pannella}, M and {Bournaud}, F and {Walter}, F and {Wang}, T and {Elbaz}, D and {Coogan}, R T},
  title = {{The [C ii] emission as a molecular gas mass tracer in galaxies at low and high redshifts}},
  journal = {\mnras},
  year = {2018},
  volume = {481},
  pages = {1976--1999},
  doi = {10.1093/mnras/sty2394},
  adsurl = {https://ui.adsabs.harvard.edu/abs/2018MNRAS.481.1976Z}
}

@INPROCEEDINGS{Hu1997,
  author = {{Hu}, E. M. and {McMahon}, R. G. and {Egami}, E.},
  title = {{Galaxies at z $>$ 4.5}},
  booktitle = {The Hubble Space Telescope and the High Redshift Universe},
  editor = {{Tanvir}, N. R. and {Arag{\'o}n-Salamanca}, A. and {Wall}, J. V.},
  publisher = {World Scientific},
  address = {Singapore},
  year = 1997,
  pages = {91},
  eprint = {astro-ph/9609188},
  archivePrefix = {arXiv},
  adsurl = {https://ui.adsabs.harvard.edu/abs/1997hsth.conf...91H}
}

@ARTICLE{HeintzWatson2020,
  author = {{Heintz}, Kasper E. and {Watson}, Darach},
  title = {{Direct Measurement of the [C I] Luminosity to Molecular Gas Mass Conversion Factor in High-redshift Star-forming Galaxies}},
  journal = {\apjl},
  year = 2020,
  volume = {889},
  pages = {L7},
  doi = {10.3847/2041-8213/ab6733},
  archivePrefix = {arXiv},
  eprint = {2001.05770},
  adsurl = {https://ui.adsabs.harvard.edu/abs/2020ApJ...889L...7H}
}

@ARTICLE{Fontana1996,
  author = {{Fontana}, A. and {Cristiani}, S. and {D'Odorico}, S. and {Giallongo}, E. and {Savaglio}, S.},
  title = {{The optical identification of a primeval galaxy at z $\gtrsim$ 4.4}},
  journal = {\mnras},
  year = 1996,
  volume = {279},
  pages = {L27},
  doi = {10.1093/mnras/279.3.l27},
  adsurl = {https://ui.adsabs.harvard.edu/abs/1996MNRAS.279L..27F}
}

@ARTICLE{Speagle2014,
  author = {{Speagle}, J. S. and {Steinhardt}, C. L. and {Capak}, P. L. and {Silverman}, J. D.},
  title = {{A Highly Consistent Framework for the Evolution of the Star-Forming Main Sequence from z $\sim$ 0-6}},
  journal = {\apjs},
  year = 2014,
  volume = {214},
  pages = {15},
  doi = {10.1088/0067-0049/214/2/15},
  adsurl = {https://ui.adsabs.harvard.edu/abs/2014ApJS..214...15S}
}

@ARTICLE{Fixsen1999,
  author = {{Fixsen}, D. J. and {Bennett}, C. L. and {Mather}, J. C.},
  title = {{COBE Far Infrared Absolute Spectrophotometer Observations of Galactic Lines}},
  journal = {\apj},
  year = 1999,
  volume = {526},
  pages = {207},
  doi = {10.1086/307962},
  adsurl = {https://ui.adsabs.harvard.edu/abs/1999ApJ...526..207F}
}

@ARTICLE{Kroupa2001,
  author = {{Kroupa}, P.},
  title = {{On the variation of the initial mass function}},
  journal = {\mnras},
  year = 2001,
  volume = {322},
  pages = {231},
  doi = {10.1046/j.1365-8711.2001.04022.x},
  archivePrefix = {arXiv},
  eprint = {astro-ph/0009005},
  adsurl = {https://ui.adsabs.harvard.edu/abs/2001MNRAS.322..231K}
}

@ARTICLE{Schoier2005,
  author = {{Sch{\"o}ier}, F. L. and {van der Tak}, F. F. S. and {van Dishoeck}, E. F. and {Black}, J. H.},
  title = {{An atomic and molecular database for analysis of submillimetre line observations}},
  journal = {\aap},
  year = 2005,
  volume = {432},
  pages = {369},
  doi = {10.1051/0004-6361:20041729},
  archivePrefix = {arXiv},
  eprint = {astro-ph/0411110},
  adsurl = {https://ui.adsabs.harvard.edu/abs/2005A&A...432..369S}
}

@ARTICLE{PoundWolfire2023,
  author = {{Pound}, Marc W. and {Wolfire}, Mark G.},
  title = {{The PhotoDissociation Region Toolbox: Software and Models for Astrophysical Analysis}},
  journal = {\aj},
  year = 2023,
  volume = {165},
  pages = {25},
  doi = {10.3847/1538-3881/ac9b1f},
  archivePrefix = {arXiv},
  eprint = {2210.08062},
  adsurl = {https://ui.adsabs.harvard.edu/abs/2023AJ....165...25P}
}

@ARTICLE{Wagg2012,
  author = {{Wagg}, J. and {Wiklind}, T. and {Carilli}, C.~L. and others},
  title = {{[C II] Line Emission in Massive Star-forming Galaxies at z = 4.7}},
  journal = {\apjl},
  year = 2012,
  volume = {752},
  eid = {L30},
  pages = {L30},
  doi = {10.1088/2041-8205/752/2/L30},
  adsurl = {https://ui.adsabs.harvard.edu/abs/2012ApJ...752L..30W}
}

@ARTICLE{Decarli2017,
  author = {{Decarli}, R. and {Walter}, F. and {Venemans}, B.~P. and {Ba{\~n}ados}, E. and {Bertoldi}, F. and {Carilli}, C. and {Fan}, X. and {Farina}, E.~P. and {Mazzucchelli}, C. and {Riechers}, D. and {Rix}, H.-W. and {Strauss}, M.~A. and {Wang}, R. and {Yang}, Y.},
  title = "{Rapidly star-forming galaxies adjacent to quasars at redshifts exceeding 6}",
  journal = {\nat}, year = 2017, volume = {545}, pages = {457},
  doi = {10.1038/nature22358},
  adsurl = {https://ui.adsabs.harvard.edu/abs/2017Natur.545..457D}
}

@ARTICLE{Valentino2020,
  author = {{Valentino}, F. and {Daddi}, E. and {Puglisi}, A. and {Magdis}, G.~E. and {Liu}, D. and {Kokorev}, V. and {Cortzen}, I. and {Madden}, S. and others},
  title = {CO emission in distant galaxies on and above the main sequence},
  journal = {\aap},
  year = {2020},
  volume = {641},
  eid = {A155},
  pages = {A155},
  doi = {10.1051/0004-6361/202038322},
  adsurl = {https://ui.adsabs.harvard.edu/abs/2020A&A...641A.155V}
}

@ARTICLE{Papadopoulos2012,
  author = {{Papadopoulos}, P. P. and {van der Werf}, P. P. and {Xilouris}, E. M. and {Isaak}, K. G. and {Gao}, Y. and {M{\"u}hle}, S.},
  title = {{The molecular gas in luminous infrared galaxies - I. CO lines, extreme physical conditions and their drivers}},
  journal = {\mnras},
  year = {2012},
  volume = {426},
  number = {4},
  pages = {2601-2629},
  doi = {10.1111/j.1365-2966.2012.21001.x},
  eprint = {1109.4176},
  adsurl = {https://ui.adsabs.harvard.edu/abs/2012MNRAS.426.2601P},
}

@ARTICLE{Vincenzo2016,
  author = {{Vincenzo}, F. and {Matteucci}, F. and {Belfiore}, F. and {Maiolino}, R.},
  title = {{Modern yields per stellar generation: the effect of the IMF}},
  journal = {\mnras},
  year = {2016},
  volume = {455},
  pages = {4183--4190},
  doi = {10.1093/mnras/stv2598},
  adsurl = {https://ui.adsabs.harvard.edu/abs/2016MNRAS.455.4183V}
}

@ARTICLE{Lehnert2020,
  author = {{Lehnert}, M. D. and {Yang}, C. and {Emonts}, B. H. C. and {Omont}, A. and {Falgarone}, E. and {Cox}, P. and {Guillard}, P.},
  title = {{Etching glass in the early Universe: Luminous HF and H$_{2}$O emission in a QSO-SMG pair at z = 4.7}},
  journal = {\aap},
  year = {2020},
  volume = {641},
  eid = {A124},
  pages = {A124},
  doi = {10.1051/0004-6361/201935755},
  adsurl = {https://ui.adsabs.harvard.edu/abs/2020A&A...641A.124L},
}

@ARTICLE{Kade2023,
  author = {{Kade}, K. and {Knudsen}, K. K. and {Vlemmings}, W. and {Stanley}, F. and {Gullberg}, B. and {K{\"o}nig}, S.},
  title = {{Exploring the environment, magnetic fields, and feedback effects of massive high-redshift galaxies with [C II]}},
  journal = {\aap},
  year = {2023},
  volume = {673},
  eid = {A116},
  pages = {A116},
  doi = {10.1051/0004-6361/202141839},
  adsurl = {https://ui.adsabs.harvard.edu/abs/2023A&A...673A.116K},
}

@ARTICLE{Aravena2024,
  author = {{Aravena}, M. and {Heintz}, K. and {Dessauges-Zavadsky}, M. and {Oesch}, P. and {Algera}, H. and {Bouwens}, R. and {da Cunha}, E. and {Dayal}, P. and {De Looze}, I. and {Ferrara}, A. and {Fudamoto}, Y. and {Gonzalez}, V. and {Graziani}, L. and {Hygate}, A. P. S. and {Inami}, H. and {Pallottini}, A. and {Schneider}, R. and {Schouws}, S. and {Sommovigo}, L. and {Topping}, M. and {van der Werf}, P. and {Palla}, M.},
  title = {{The ALMA Reionization Era Bright Emission Line Survey: The molecular gas content of galaxies at z $\sim$ 7}},
  journal = {\aap},
  year = {2024},
  volume = {682},
  eid = {A24},
  pages = {A24},
  doi = {10.1051/0004-6361/202347281},
  adsurl = {https://ui.adsabs.harvard.edu/abs/2024A&A...682A..24A},
}

@ARTICLE{Bakx2024,
  author = {{Bakx}, T. J. L. C. and {Algera}, H. S. B. and {Venemans}, B. and {Sommovigo}, L. and {Fujimoto}, S. and {Carniani}, S. and {Hagimoto}, M. and {Hashimoto}, T. and {Inoue}, A. K. and {Salak}, D. and {Serjeant}, S. and {Vallini}, L. and {Eales}, S. and {Ferrara}, A. and {Fudamoto}, Y. and {Imamura}, C. and {Inoue}, S. and {Knudsen}, K. K. and {Matsuo}, H. and {Sugahara}, Y. and {Tamura}, Y. and {Taniguchi}, A. and {Yamanaka}, S.},
  title = {{Gas conditions of a star-formation selected sample in the first billion years}},
  journal = {\mnras},
  year = {2024},
  volume = {532},
  pages = {2270-2288},
  doi = {10.1093/mnras/stae1613},
  adsurl = {https://ui.adsabs.harvard.edu/abs/2024MNRAS.532.2270B},
}

@ARTICLE{Shen2007,
  author = {{Shen}, Y. and {Strauss}, M. A. and {Oguri}, M. and {Hennawi}, J. F. and {Fan}, X. and {Richards}, G. T. and {Hall}, P. B. and {Gunn}, J. E. and {Schneider}, D. P. and {Szalay}, A. S. and {Thakar}, A. R. and {Vanden Berk}, D. E. and {Anderson}, S. F. and {Bahcall}, N. A. and {Connolly}, A. J. and {Knapp}, G. R.},
  title = {{Clustering of High-Redshift (z$\geq$ 2.9) Quasars from the Sloan Digital Sky Survey}},
  journal = {\aj},
  year = {2007},
  volume = {133},
  pages = {2222-2241},
  doi = {10.1086/513517},
  adsurl = {https://ui.adsabs.harvard.edu/abs/2007AJ....133.2222S},
}

@ARTICLE{Navarro1997,
  author = {{Navarro}, J. F. and {Frenk}, C. S. and {White}, S. D. M.},
  title = {{A Universal Density Profile from Hierarchical Clustering}},
  journal = {\apj},
  year = {1997},
  volume = {490},
  pages = {493-508},
  doi = {10.1086/304888},
  adsurl = {https://ui.adsabs.harvard.edu/abs/1997ApJ...490..493N},
}

@ARTICLE{Madau2014,
       author = {{Madau}, Piero and {Dickinson}, Mark},
        title = "{Cosmic Star-Formation History}",
      journal = {\araa},
         year = 2014,
        month = aug,
       volume = {52},
        pages = {415-486},
          doi = {10.1146/annurev-astro-081811-125615},
archivePrefix = {arXiv},
       eprint = {1403.0007},
 primaryClass = {astro-ph.CO},
       adsurl = {https://ui.adsabs.harvard.edu/abs/2014ARA&A..52..415M}
}

@ARTICLE{Lee2019,
       author = {{Lee}, Minju M. and {Nagao}, Tohru and {De Breuck}, Carlos and {Carniani}, Stefano and {Cresci}, Giovanni and {Hatsukade}, Bunyo and {Kawabe}, Ryohei and {Kohno}, Kotaro and {Maiolino}, Roberto and {Mannucci}, Filippo and {Marconi}, Alessandro and {Nakanishi}, Kouichiro and {Saito}, Toshiki and {Tamura}, Yoichi and {Troncoso}, Paulina and {Umehata}, Hideki and {Yun}, Min},
        title = "{First [N II]122 {\ensuremath{\mu}}m Line Detection in a QSO-SMG Pair BRI 1202-0725 at z = 4.69}",
      journal = {\apjl},
         year = 2019,
        month = oct,
       volume = {883},
       number = {2},
          eid = {L29},
        pages = {L29},
          doi = {10.3847/2041-8213/ab412e},
archivePrefix = {arXiv},
       eprint = {1909.02284},
 primaryClass = {astro-ph.GA},
       adsurl = {https://ui.adsabs.harvard.edu/abs/2019ApJ...883L..29L}
}

@ARTICLE{Yu2021,
       author = {{Yu}, Xiaodi and {Li}, Jiang-Tao and {Qu}, Zhijie and {Roederer}, Ian U. and {Bregman}, Joel N. and {Fan}, Xiaohui and {Fang}, Taotao and {Johnson}, Sean D. and {Wang}, Feige and {Yang}, Jinyi},
        title = "{Probing the He II re-Ionization ERa via Absorbing C IV Historical Yield (HIERACHY) I: A strong outflow from a z {\ensuremath{\sim}} 4.7 quasar}",
      journal = {\mnras},
         year = 2021,
        month = aug,
       volume = {505},
       number = {3},
        pages = {4444-4455},
          doi = {10.1093/mnras/stab1614},
archivePrefix = {arXiv},
       eprint = {2105.13498},
 primaryClass = {astro-ph.GA},
       adsurl = {https://ui.adsabs.harvard.edu/abs/2021MNRAS.505.4444Y}
}

@ARTICLE{Scoville2016,
       author = {{Scoville}, N. and {Sheth}, K. and {Aussel}, H. and {Vanden Bout}, P. and {Capak}, P. and {Bongiorno}, A. and {Casey}, C.~M. and {Murchikova}, L. and {Koda}, J. and {{\'A}lvarez-M{\'a}rquez}, J. and {Lee}, N. and {Laigle}, C. and {McCracken}, H.~J. and {Ilbert}, O. and {Pope}, A. and {Sanders}, D. and {Chu}, J. and {Toft}, S. and {Ivison}, R.~J. and {Manohar}, S.},
        title = "{ISM Masses and the Star formation Law at Z = 1 to 6: ALMA Observations of Dust Continuum in 145 Galaxies in the COSMOS Survey Field}",
      journal = {\apj},
         year = 2016,
        month = apr,
       volume = {820},
       number = {2},
          eid = {83},
        pages = {83},
          doi = {10.3847/0004-637X/820/2/83},
archivePrefix = {arXiv},
       eprint = {1511.05149},
 primaryClass = {astro-ph.GA},
       adsurl = {https://ui.adsabs.harvard.edu/abs/2016ApJ...820...83S}
}

@ARTICLE{Bohlin1978,
       author = {{Bohlin}, R.~C. and {Savage}, B.~D. and {Drake}, J.~F.},
        title = "{A survey of interstellar H I from Lalpha absorption measurements. II.}",
      journal = {\apj},
         year = 1978,
        month = aug,
       volume = {224},
        pages = {132-142},
          doi = {10.1086/156357},
       adsurl = {https://ui.adsabs.harvard.edu/abs/1978ApJ...224..132B}
}

@ARTICLE{Villanueva2024,
       author = {{Villanueva}, V. and {Herrera-Camus}, R. and {Gonz{\'a}lez-L{\'o}pez}, J. and {Aravena}, M. and {Assef}, R.~J. and {Baeza-Garay}, M. and {Barcos-Mu{\~n}oz}, L. and {Bovino}, S. and {Bowler}, R.~A.~A. and {da Cunha}, E. and {De Looze}, I. and {Diaz-Santos}, T. and {Ferrara}, A. and {F{\"o}rster Schreiber}, N.~M. and {Algera}, H. and {Ikeda}, R. and {Killi}, M. and {Mitsuhashi}, I. and {Naab}, T. and {Relano}, M. and {Spilker}, J. and {Solimano}, M. and {Palla}, M. and {Price}, S.~H. and {Posses}, A. and {Tadaki}, K. and {Telikova}, K. and {{\"U}bler}, H.},
        title = "{The ALMA-CRISTAL survey: Dust temperature and physical conditions of the interstellar medium in a typical galaxy at z = 5.66}",
      journal = {\aap},
         year = 2024,
        month = nov,
       volume = {691},
          eid = {A133},
        pages = {A133},
          doi = {10.1051/0004-6361/202451490},
archivePrefix = {arXiv},
       eprint = {2407.09681},
 primaryClass = {astro-ph.GA},
       adsurl = {https://ui.adsabs.harvard.edu/abs/2024A&A...691A.133V}
}

@ARTICLE{Algera2024,
       author = {{Algera}, Hiddo S.~B. and {Inami}, Hanae and {De Looze}, Ilse and {Ferrara}, Andrea and {Hirashita}, Hiroyuki and {Aravena}, Manuel and {Bakx}, Tom and {Bouwens}, Rychard and {Bowler}, Rebecca A.~A. and {Da Cunha}, Elisabete and {Dayal}, Pratika and {Fudamoto}, Yoshinobu and {Hodge}, Jacqueline and {Hygate}, Alexander and {van Leeuwen}, Ivana and {Nanayakkara}, Themiya and {Palla}, Marco and {Pallottini}, Andrea and {Rowland}, Lucie and {Smit}, Renske and {Sommovigo}, Laura and {Stefanon}, Mauro and {Vijayan}, Aswin P. and {van der Werf}, Paul},
        title = "{Accurate simultaneous constraints on the dust mass, temperature, and emissivity index of a galaxy at redshift 7.31}",
      journal = {\mnras},
         year = 2024,
        month = sep,
       volume = {533},
       number = {3},
        pages = {3098-3113},
          doi = {10.1093/mnras/stae1994},
archivePrefix = {arXiv},
       eprint = {2408.02267},
 primaryClass = {astro-ph.GA},
       adsurl = {https://ui.adsabs.harvard.edu/abs/2024MNRAS.533.3098A}
}

@ARTICLE{Faisst2020,
       author = {{Faisst}, Andreas L. and {Fudamoto}, Yoshinobu and {Oesch}, Pascal A. and {Scoville}, Nick and {Riechers}, Dominik A. and {Pavesi}, Riccardo and {Capak}, Peter},
        title = "{ALMA characterizes the dust temperature of z {\ensuremath{\sim}} 5.5 star-forming galaxies}",
      journal = {\mnras},
         year = 2020,
        month = nov,
       volume = {498},
       number = {3},
        pages = {4192-4204},
          doi = {10.1093/mnras/staa2545},
archivePrefix = {arXiv},
       eprint = {2005.07716},
 primaryClass = {astro-ph.GA},
       adsurl = {https://ui.adsabs.harvard.edu/abs/2020MNRAS.498.4192F}
}

@ARTICLE{Demyk2017,
       author = {{Demyk}, K. and {Meny}, C. and {Lu}, X.-H. and {Papatheodorou}, G. and {Toplis}, M.~J. and {Leroux}, H. and {Depecker}, C. and {Brubach}, J.-B. and {Roy}, P. and {Nayral}, C. and {Ojo}, W.-S. and {Delpech}, F. and {Paradis}, D. and {Gromov}, V.},
        title = "{Low temperature MIR to submillimeter mass absorption coefficient of interstellar dust analogues. I. Mg-rich glassy silicates}",
      journal = {\aap},
         year = 2017,
        month = apr,
       volume = {600},
          eid = {A123},
        pages = {A123},
          doi = {10.1051/0004-6361/201629711},
archivePrefix = {arXiv},
       eprint = {1701.07225},
 primaryClass = {astro-ph.GA},
       adsurl = {https://ui.adsabs.harvard.edu/abs/2017A&A...600A.123D}
}

@ARTICLE{Zhang2016,
       author = {{Zhang}, Zhi-Yu and {Papadopoulos}, Padelis P. and {Ivison}, R.~J. and {Galametz}, Maud and {Smith}, M.~W.~L. and {Xilouris}, Emmanuel M.},
        title = "{Gone with the heat: a fundamental constraint on the imaging of dust and molecular gas in the early Universe}",
      journal = {Royal Society Open Science},
         year = 2016,
       volume = {3},
        pages = {160025},
          doi = {10.1098/rsos.160025},
       adsurl = {https://ui.adsabs.harvard.edu/abs/2016RSOS....360025Z},
}

@ARTICLE{Wang2013,
       author = {{Wang}, Ran and {Wagg}, Jeff and {Carilli}, Chris L. and {Walter}, Fabian and {Lentati}, Lindley and {Fan}, Xiaohui and {Riechers}, Dominik A. and {Bertoldi}, Frank and {Narayanan}, Desika and {Strauss}, Michael A. and {Cox}, Pierre and {Omont}, Alain and {Menten}, Karl M. and {Knudsen}, Kirsten K. and {Neri}, Roberto and {Jiang}, Linhua},
        title = "{Star Formation and Gas Kinematics of Quasar Host Galaxies at z \raisebox{-0.5ex}\textasciitilde 6: New Insights from ALMA}",
      journal = {\apj},
         year = 2013,
        month = aug,
       volume = {773},
       number = {1},
          eid = {44},
        pages = {44},
          doi = {10.1088/0004-637X/773/1/44},
archivePrefix = {arXiv},
       eprint = {1302.4154},
 primaryClass = {astro-ph.CO},
       adsurl = {https://ui.adsabs.harvard.edu/abs/2013ApJ...773...44W}
}

@article{Wang2026,
  author = {Wang, Feige and Champagne, Jaclyn B. and Huang, Jiamu and Yang, Jinyi and Hennawi, Joseph F. and Fan, Xiaohui and Zhang, Haowen and Costa, Tiago and Decarli, Roberto and Habouzit, Melanie and Sun, Fengwu and Ba{\~n}ados, Eduardo and Jin, Xiangyu and Kakiichi, Koki and Meyer, Romain A. and Wu, Yunjing and Belladitta, Silvia and Blecha, Laura and Bosman, Sarah E. I. and Cai, Zheng and Connor, Thomas and Davies, Frederick B. and Eilers, Anna-Christina and Haiman, Zolt{\'a}n and Jun, Hyunsung D. and Li, Mingyu and Li, Zihao and Liu 刘, Weizhe 伟哲 and Lupi, Alessandro and Lyu 吕, Jianwei 建伟 and Mazzucchelli, Chiara and Onoue, Masafusa and Pizzati, Elia and Pudoka, Maria and Rojas-Ruiz, Sof{\'i}a and Schindler, Jan-Torge and Shen, Yue and Tee, Wei Leong and Trakhtenbrot, Benny and Trebitsch, Maxime and Vestergaard, Marianne and Volonteri, Marta and Walter, Fabian and Zhang, Huanian and Zou, Siwei},
  title = {ASPIRE: The Environments and Dark Matter Halos of Luminous Quasars in the Epoch of Reionization},
  journal = {\apj},
  year = {2026},
  volume = {1006},
  pages = {39},
  doi = {10.3847/1538-4357/ae7bfa},
  eprint = {2602.04979},
  archivePrefix = {arXiv},
  primaryClass = {astro-ph.GA}
}

@article{Decarli2024,
  author = {Decarli, Roberto and Loiacono, Federica and Farina, Emanuele Paolo and Dotti, Massimo and Lupi, Alessandro and Meyer, Romain A. and Mignoli, Marco and Pensabene, Antonio and Strauss, Michael A. and Venemans, Bram and Yang, Jinyi and Walter, Fabian and Wolf, Julien and Ba{\~n}ados, Eduardo and Blecha, Laura and Bosman, Sarah and Carilli, Chris L. and Comastri, Andrea and Connor, Thomas and Costa, Tiago and Eilers, Anna-Christina and Fan, Xiaohui and Gilli, Roberto and Jun, Hyunsung D. and Liu, Weizhe and Marshall, Madeline A. and Mazzucchelli, Chiara and Neeleman, Marcel and Onoue, Masafusa and Overzier, Roderik and Pudoka, Maria Anne and Riechers, Dominik A. and Rix, Hans-Walter and Schindler, Jan-Torge and Trakhtenbrot, Benny and Trebitsch, Maxime and Vestergaard, Marianne and Volonteri, Marta and Wang, Feige and Zhang, Huanian and Zou, Siwei},
  title = {A quasar-galaxy merger at $z\sim 6.2$: rapid host growth via accretion of two massive satellite galaxies},
  journal = {\aap},
  year = {2024},
  volume = {689},
  pages = {A219},
  doi = {10.1051/0004-6361/202449239},
  eprint = {2406.06697},
  archivePrefix = {arXiv},
  primaryClass = {astro-ph.GA}
}

@ARTICLE{Omont1996,
       author = {{Omont}, Alain and {Petitjean}, Patrick and {Guilloteau}, St{\'e}phane and {McMahon}, Richard G. and {Solomon}, P.~M. and {P{\'e}contal}, Emmanuel},
        title = "{Molecular gas and dust around a radio-quiet quasar at redshift 4.69}",
      journal = {\nat},
         year = 1996,
        month = aug,
       volume = {382},
       number = {6590},
        pages = {428-431},
          doi = {10.1038/382428a0},
archivePrefix = {arXiv},
       eprint = {astro-ph/9608006},
       adsurl = {https://ui.adsabs.harvard.edu/abs/1996Natur.382..428O},
}

\begin{appendix}
\global\dbltextfloatsep=8pt\relax\global\textfloatsep=8pt\relax\setcounter{dbltopnumber}{3}\renewcommand{\dbltopfraction}{0.95}\renewcommand{\textfraction}{0.05}\renewcommand{\topfraction}{0.95}

\section{Spectral extraction and positions}\label{app:spectra}

\begin{figure}[t]
\includegraphics[width=86mm]{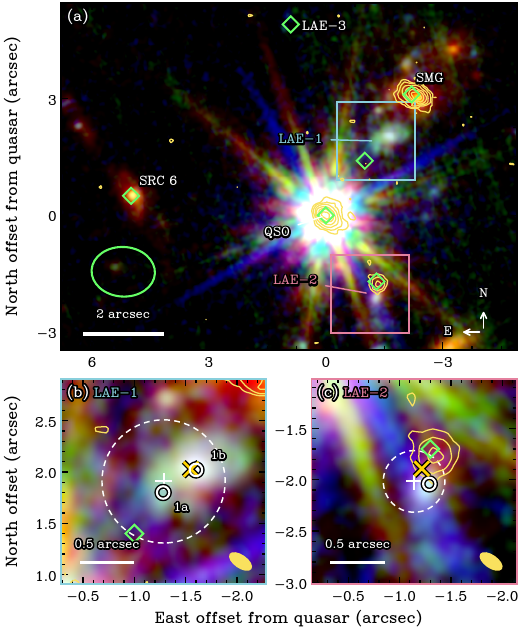}
\caption{HST F775W (blue), JWST F200W (green) and F444W (red) composites of (a) the BR1202$-$0725 group and of (b) LAE\mbox{-}1 and (c) LAE\mbox{-}2 (boxes in a), registered on the quasar position of \citet{Zamora2025}; offsets are from the quasar. Green diamonds: fixed extraction positions relative to the model quasar (green ellipse: Band~3 beam, $1.63\arcsec\times1.27\arcsec$, PA $87\degr$); contours: ALMA Band~7 continuum \citep{Lelli2026} at 3, 6, 12, 24, 48 and 96 times 36\,$\mu$Jy\,beam$^{-1}$ (beam: yellow ellipses); white circles: stellar peaks (1a and 1b in b); yellow crosses: light-weighted JWST centroids; plus signs and dashed circles: the 0.6\arcsec\ (LAE\mbox{-}1) and 0.3\arcsec\ (LAE\mbox{-}2) radius apertures of \citet{Zamora2025}.}\label{fig:positions}
\end{figure}

\emph{Visibility fits.} The calibrated visibilities, without continuum subtraction, were fitted channel by channel with \mbox{UVMULTIFIT} \citep{MartiVidal2014}, with six point sources at fixed positions and free flux densities, which the visibilities separate through their position-dependent phases; the corrupted first channel of each of the 16 spectral windows was excluded. Continuum and lines were then fitted to the extracted spectra. In line-free channels, the fluxes scatter by 0.70 times the uncertainties returned by \mbox{UVMULTIFIT} for all components and tunings, without significant adjacent-channel correlation; we scale the errors by this factor. Line profiles were fitted to the unbinned spectra, and the flux errors include the covariance of the profile parameters but not the 5\% flux-calibration uncertainty. For the quasar host and the SMG, our CO(5\mbox{--}4) fluxes, $1.581\pm0.023$ and $2.360\pm0.043$\,\Jykms, agree with the $1.6\pm0.2$ and $2.6\pm0.4$\,\Jykms\ of \citet{Salome2012}, and their \ci(1\mbox{--}0) fluxes are $0.296\pm0.016$ and $0.405\pm0.028$\,\Jykms.

\emph{Positions and separations.} The six positions are phase-center offsets (arcsec; Fig.~\ref{fig:positions}): the quasar ($+0.2$, $-0.5$), the SMG ($-2.0$, $+2.6$), LAE\mbox{-}1 ($-0.8$, $+0.9$) at the narrow \cii\ peak of \citet{Carilli2013}, LAE\mbox{-}2 ($-1.1$, $-2.2$) at its dust peak, 0.2--0.35\arcsec\ north of the stellar light, and the \lya\ emitter LAE\mbox{-}3 ($+1.1$, $+4.4$) and the faint continuum source SRC~6 ($+5.2$, $0.0$) at the MUSE and dust-continuum positions of \citet{Drake2020}. These are fixed inputs, not CO centroids; only the quasar position was set by a continuum-visibility fit. The images in Fig.~\ref{fig:positions} are registered on their quasar cores, a relative overlay rather than an absolute astrometric registration. We adopt a quasar--LAE\mbox{-}2 separation of 2.19\arcsec\ (14.5\,kpc), measured between the Band~3 quasar centroid and the \cii\ kinematic center of \citet{Lelli2026}. The LAE\mbox{-}1 position, hereafter the extraction position, is 1.7\arcsec\ from the quasar, whereas the stellar peaks LAE\mbox{-}1a and LAE\mbox{-}1b, at ($-1.08$, $+1.30$) and ($-1.40$, $+1.52$), lie at 2.2\arcsec\ and 2.6\arcsec, and the 3\arcsec\ quoted in earlier work refers to the \lya\ peak. The extraction position is 0.49\arcsec\ (3.2\,kpc) from LAE\mbox{-}1a and 0.84\arcsec\ (5.5\,kpc) from the light-weighted centroid of the pair (Fig.~\ref{fig:positions}b), and the fit at this position recovers 0.83 and 0.59 of the flux of a point source at LAE\mbox{-}1a and LAE\mbox{-}1b.

\begin{figure}[t]
\includegraphics[width=85mm]{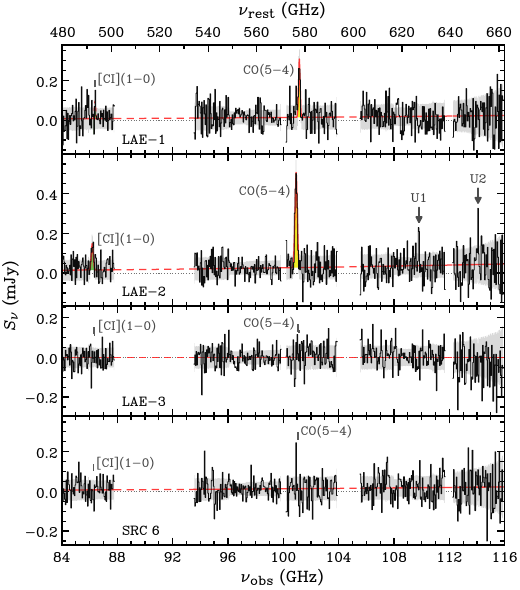}
\caption{Full-band spectra from the visibility model in 50\,MHz bins. Red: power-law continuum plus CO(5\mbox{--}4) and \ci(1\mbox{--}0) (continuum only for LAE\mbox{-}3 and SRC~6), dashed where undetected and across the tuning gaps; gray: $\pm1\sigma$; yellow and green: CO(5\mbox{--}4) and \ci(1\mbox{--}0) channels within one FWHM of the line center. Top axis: rest frequency at the CO redshift of LAE\mbox{-}2. Ticks: expected frequencies of undetected lines, for LAE\mbox{-}3 at $z_{\rm Ly\alpha}=4.7019$ (bars: up to 400\,\kms\ blueward) and for SRC~6 at $z=4.7025$--4.7035 \citep{Drake2020}. U1, U2: tentative unidentified features.}\label{fig:fullspectra}
\end{figure}

\emph{Continuum.} We fitted a power law, $S_\nu\propto\nu^{3.7}$, to the line-free channels of each faint component. LAE\mbox{-}2 has $S_{100}=26.4\pm2.3$\,$\mu$Jy, unchanged when individual windows or tunings are omitted. The LAE\mbox{-}1 position has $13.6\pm2.4$\,$\mu$Jy, but its four tunings scatter by 5.2\,$\mu$Jy, so we use this candidate only as an upper limit, $S_{100}<31$\,$\mu$Jy (App.~\ref{app:dustpdr}).

\emph{LAE\mbox{-}3 and SRC~6.} Neither source shows a line (limits of 0.05\,\Jykms\ for CO(5\mbox{--}4) and 0.043\,\Jykms\ for \ci(1\mbox{--}0) at a FWHM of 300\,\kms). LAE\mbox{-}3 is also undetected in the continuum ($S_{100}<7$\,$\mu$Jy, and at 223 and 426\,GHz), whereas SRC~6 is detected at 100\,GHz ($13.2\pm2.3$\,$\mu$Jy) and at 223\,GHz ($92\pm24$\,$\mu$Jy in a 0.5\arcsec\ aperture), but not at 426\,GHz. Its CO(5\mbox{--}4) limit lies at least $4\sigma$ below the value expected for the CO-to-continuum ratio of LAE\mbox{-}2, so SRC~6 either is not at $z\approx4.70$ or has weaker CO per unit dust.

\emph{Line census and unidentified features.} A blind matched-filter search of the four faint components finds four features beyond $\pm4\sigma$ besides CO(5\mbox{--}4) in LAE\mbox{-}2 and the signal at the LAE\mbox{-}1 position, where Gaussian noise predicts fewer than one. Scaled errors underestimated by about 10\% would explain this excess, and we use such an inflation as a sensitivity test, in which \ci\ in LAE\mbox{-}2 is at $3.5\sigma$ and the CO(5\mbox{--}4) signal at the LAE\mbox{-}1 position at $4.5\sigma$. At the LAE\mbox{-}1 position, the spectrum shows no line other than the CO(5\mbox{--}4) signal at 5.0$\sigma$. In LAE\mbox{-}2, two further features, U1 at 109.81\,GHz and U2 at 114.08\,GHz (rest-frame 627 and 651\,GHz), have $0.077\pm0.021$ and $0.109\pm0.032$\,\Jykms. Both lie at spectral-window junctions, with positive contributions in both adjacent windows and no corresponding feature in the other five components. U2 lies within 330\,\kms\ of the N$_2$H$^+$(7\mbox{--}6) transition. We treat both as tentative features, as noise, calibration, or modeling residuals could also produce them.

\emph{The signal at the LAE\mbox{-}1 position.} To locate the emission, we repeated the visibility fit with the LAE\mbox{-}1 component at LAE\mbox{-}1a and at LAE\mbox{-}1b; the number of free parameters is unchanged, so the $\chi^2$ values can be compared directly. With the scaled errors, $\chi^2$ over the 33 CO(5\mbox{--}4) channels within one FWHM of the line center is higher by 23 at LAE\mbox{-}1a and by 49 at LAE\mbox{-}1b than at the extraction position. In simulations with the measured noise, a point source at the extraction position gives increases of $13\pm10$ and $27\pm14$ (mean and standard deviation), consistent with those observed, whereas a compact source at the LAE\mbox{-}1a or LAE\mbox{-}1b stellar peak (Fig.~\ref{fig:positions}) alone gives such increases with a probability of at most $2\times10^{-4}$ ($1.5\times10^{-3}$ with the 10\% inflation). The fixed quasar position lies 0.09--0.14\arcsec\ from its image-plane centroids, which could transfer 3--5\% of the quasar flux to the LAE\mbox{-}1 component. A joint fit of the spectra at the three positions with scaled copies of the quasar and SMG spectra, however, gives a quasar copy with $-0.2\pm1.5$\% of the quasar CO flux and a negative SMG copy, whereas the source remains at $4.5\sigma$, and at $3.4\sigma$ when a velocity gradient across the quasar host \citep{Lelli2026} is also left free. The \cii\ data of \citet{Lelli2026} point the same way: they show no compact source at the extraction position ($<0.39$\,\Jykms\ within 0.3\arcsec), but a simultaneous point-source fit at the quasar, SMG, LAE\mbox{-}2 and extraction positions on the cube convolved to the Band~3 beam shows a line twice as broad as the CO (FWHM $843\pm123$\,\kms\ for a free Gaussian) and gives $2.55\pm0.37$\,\Jykms\ for a Gaussian with the CO centroid and width. The resulting \cii-to-CO(5\mbox{--}4) flux ratio, $21\pm5$, is comparable to the total-line ratio of LAE\mbox{-}2 (18), and the whole \cii\ line gives $39\pm9$; both exceed the ratios of the quasar host (7) and the SMG (8). We conclude that the line most likely arises from extended gas at the extraction position, co-spatial with the narrow \cii\ peak of \citet{Carilli2013}, rather than from either stellar peak. Its velocity lies between those of the quasar and LAE\mbox{-}1a (App.~\ref{app:positions}), so it may be tidal gas associated with LAE\mbox{-}1.

\section{Kinematics and dynamical masses}\label{app:positions}

\emph{Kinematics.} In velocity, the CO line at the LAE\mbox{-}1 position is closer to the narrow \cii\ emission than to either stellar peak: its redshift is $+93\pm35$\,\kms\ from the narrow \cii\ \citep{Carilli2013} and $-1\pm44$\,\kms\ from the tentative \nii\,205\,$\mu$m feature of \citet{Decarli2014}, which later observations did not detect \citep{Pavesi2016,Lu2017}, but $-207\pm35$\,\kms\ from H$\alpha$ of LAE\mbox{-}1a and $-802\pm35$\,\kms\ from LAE\mbox{-}1b \citep{Zamora2025}. For LAE\mbox{-}2, the CO redshift agrees with the \cii\ redshift of \citet{Lelli2026} within its rounding ($\pm26$\,\kms) and with the CO(2\mbox{--}1) window center of \citet{Jones2016} within 25\,\kms, whereas H$\alpha$ is 167\,\kms\ bluer \citep{Zamora2025}, which explains the 650\,\kms\ offset quoted in earlier work. For an assumed $10^{12.5}$--$10^{13}\,\Msun$ halo \citep{Shen2007}, the $+820$\,\kms\ of LAE\mbox{-}2 relative to the quasar is 1.4--2.0 virial velocities, below the escape speed of a \citet{Navarro1997} halo with a concentration of 3--5 at 0.12--0.18 virial radii (2.7--2.9 virial velocities), so LAE\mbox{-}2 may be bound; the line-of-sight velocity alone does not decide this.

\emph{Dynamical masses.} The dynamical mass of LAE\mbox{-}2 bounds its gas mass and thus $\aco$. If the CO traces a rotating disk, $M_{\rm dyn}(<R)=v_{\rm circ}^{2}R/G=2.33\times10^{5}\,v_{\rm circ}^{2}R$ in $\Msun$, with $v_{\rm circ}=0.75\,{\rm FWHM}/\sin i$ in \kms\ and $R$ in kpc \citep{Wang2013}. The gas radius is not measured: the compact dust (half-light radius $\approx1$\,kpc) is smaller than the extended \cii\ emission \citep{Lelli2026} and the LVG equivalent radius (1.6\,kpc). We therefore adopt half the circularized continuum FWHM of \citet{Lu2017}, with and without beam deconvolution, $R=2.0$--3.6\,kpc, which gives $M_{\rm dyn}=(8.5$--$61)\times10^{10}\,\Msun$ for $i=90$--30\degr. If this radius encloses all the gas and stars, the largest $\aco$ allowed by the dynamical mass, $(M_{\rm dyn}-\Mstar)/\Lp_{\rm CO(1-0)}$ for the CMB-corrected luminosity, is 1.0--2.0 edge-on and 4.8--8.9 at $i=30\degr$; it is an upper bound, because it neglects dark matter. The \ci-based mass ($\aco=1.4$) is then compatible with the dynamical mass for inclinations up to about 60\degr\ at $R=2$\,kpc. In contrast, a Galactic $\aco=4.4$ requires an inclination below 31--44\degr\ (14--28\% of random orientations), depending on the radius. If only half of the gas lies within $R$, as assumed for the shielding column (App.~\ref{app:dustpdr}), these limits relax to 45--73\degr\ (30--70\%), illustrating the dependence on gas distribution.

\section{Non-LTE excitation analysis}\label{app:radex}

\emph{Data.} The CO(2\mbox{--}1) flux of \citet{Jones2016}, $0.19\pm0.03$\,\Jykms\ over $+450$ to $+1240$\,\kms, misses part of the CO(5\mbox{--}4) profile, so the model CO(2\mbox{--}1) flux is multiplied by the window fraction $f_{\rm win}=0.894^{+0.035}_{-0.040}$ of a Gaussian with the CO(5\mbox{--}4) centroid and width. The quoted $r_{52}$ uses the measured flux divided by $f_{\rm win}$, which assumes that CO(2\mbox{--}1) follows the CO(5\mbox{--}4) profile; without this correction, $r_{52}$ would be 0.25. Flux-scale terms of 5\% are added in quadrature to both lines. The CO(12\mbox{--}11) limit of \citet{Lee2021} (Table~\ref{tab:anc}) does not constrain the fit.

\emph{LVG modeling of CO excitation.} As in \citet{Yang2017}, we ran RADEX \citep{vanderTak2007} in the expanding-sphere (LVG) geometry with the LAMDA CO rates \citep{Schoier2005}, $T_{\rm CMB}=15.56$\,K and $\Delta v=571$\,\kms. The free parameters are $\log n_{\rm H_2}\in[2,7]$, $\log T_{\rm kin}\in[\log T_{\rm CMB},3]$, $\log(N_{\rm CO}/\Delta v)\in[15.5,19.5]$ (cm$^{-2}$\,km$^{-1}$\,s) and the source solid angle $\Omega$, log-uniform between equivalent radii of 0.02 and 3\arcsec. Draws with CO optical depths above 100 are rejected, and for a CO abundance $x_{\rm CO}=10^{-4}$ the velocity gradient is restricted to 0.1--1000\,\kms\,pc$^{-1}$. We used emcee \citep[48 walkers, 7000 steps after 800 burn-in steps;][]{ForemanMackey2013}. The posterior medians are $\log(n_{\rm H_2}/{\rm cm^{-3}})=2.75^{+0.71}_{-0.50}$, $T_{\rm kin}=126^{+308}_{-90}$\,K, $\log(N_{\rm CO}/\Delta v)=16.39^{+0.72}_{-0.60}$ and an equivalent radius of $R_{\rm eq}=1.6^{+0.8}_{-0.4}$\,kpc, with $T_{\rm ex}(5$--$4)=19.5$\,K and an observed $\Lp_{\rm CO(1-0)}=(4.1\pm1.1)\times10^{10}$\,\Lpunit. The CMB contrasts, the observed fractions of the intrinsic line fluxes, are 0.63, 0.36 and 0.74 for CO(1\mbox{--}0), CO(5\mbox{--}4) and \ci, so the CMB-corrected $\Lp_{\rm CO(1-0)}$ and $r_{52}$ are $6.7^{+3.6}_{-2.6}\times10^{10}$\,\Lpunit\ and 0.33. For the same $\aco$, LAE\mbox{-}2 is thus $4.0^{+2.5}_{-1.6}$ times fainter in CO(5\mbox{--}4) per unit gas mass than optically thick, thermalized 40\,K gas (CMB-corrected CO(5\mbox{--}4)-to-CO(1\mbox{--}0) ratio of 0.74, CO(5\mbox{--}4) contrast 0.80), so a CO(5\mbox{--}4) survey of fixed depth needs four times more gas to detect such a reservoir.

\emph{Constraints and degeneracies.} The low $r_{52}$ requires sub-thermal CO(5\mbox{--}4), but not a strict density limit. Density and temperature are anti-correlated: a solution with $10^{4}$\,cm$^{-3}$ and 30\,K also fits, so $n_{\rm H_2}\lesssim10^{3.5}$\,cm$^{-3}$ is the 84th percentile of the posterior, not a physical limit. $T_{\rm kin}$ is weakly constrained and $N_{\rm CO}/\Delta v$ is bounded from above. Once $T_{\rm ex}$ is fixed, the absolute flux sets $\Omega$, and $R_{\rm eq}$ exceeds the half-light radius of the compact dust ($\approx1$\,kpc), as expected if the line emission extends beyond the dusty core (App.~\ref{app:dustpdr}). A uniform sphere or a slab instead of the expanding sphere would change $n_{\rm H_2}$ and $N_{\rm CO}/\Delta v$ by a few tenths of a dex. The implied velocity gradient, $K_{\rm vir}=16^{+65}_{-14}$ times the virial value \citep{Papadopoulos2012}, suggests gas that is not self-gravitating, as in turbulent or tidally stirred disks, although virialized gas ($K_{\rm vir}\leq3$) remains allowed; the value scales with the assumed abundance.

\emph{\ci\ and the CMB correction.} With the same $n_{\rm H_2}$, $T_{\rm kin}$ and $\Omega$, we solved for the \ci\ column in 600 posterior samples. The line is optically thin ($\tau_{\rm CI}\approx0.04$), with $T_{\rm ex}({\rm CI})=36^{+33}_{-12}$\,K and $N_{\rm CI}/N_{\rm CO}=0.52^{+0.65}_{-0.33}$, so an enhanced \ci/CO ratio is not required. We adopt $\alpha_{\rm CI}=17\,\Msun\,$(\Lpunit)$^{-1}$, including helium \citep{Dunne2022}, and apply it to the \ci\ luminosity divided by its contrast, 0.74. This gives $\Mmol=9.2^{+8.8}_{-4.3}\times10^{10}\,\Msun$ ($6.4\times10^{10}\,\Msun$ without the correction), equivalent to $\aco=1.4$ for the corrected and 2.2 for the observed CO(1\mbox{--}0) luminosity. Its uncertainty combines those of the \ci\ flux and excitation, with 0.11\,dex for the scatter in $\alpha_{\rm CI}$ and 0.20\,dex for metallicity. The metallicity-dependent $\alpha_{\rm CI}$ of \citet{HeintzWatson2020} gives the same mass within 12\% at the near-solar metallicity of LAE\mbox{-}2.

\emph{Fixed-gas test.} The CMB also pumps the CO levels. To separate this effect from the state of the gas, we reran RADEX for 5000 posterior draws at $T_{\rm CMB}=2.73$\,K with all fitted parameters fixed. For CO(5\mbox{--}4), the weaker pumping and the higher contrast cancel, so the observed luminosity of the same gas at $z=0$ is unchanged (ratio $0.98^{+0.21}_{-0.07}$), whereas CO(2\mbox{--}1) brightens by a factor of 1.5. Against the $z=0$ background, the same gas would therefore have $r_{52}=0.15^{+0.06}_{-0.04}$, below the BzK disks at $z\approx1.5$ (0.30) and the ASPECS stacked spectra at $z\approx1.2$ (0.23), close to the value of the Galactic center and twice that of the inner Galaxy (Fig.~\ref{fig:sled}b; \citealt{Fixsen1999}); the CMB-corrected ratio at $z=4.7$ is 0.33, so the CMB contrast lowers the observed ratio to 0.22, but the same gas at $z=0$ would show a lower ratio still. For \ci, the corrected luminosity and the $z=0$ luminosity of the same gas agree within 4\%, which justifies applying the $z\approx0$ calibration of $\alpha_{\rm CI}$ to the corrected line.

\emph{Two-component test.} Adding a warmer, thermalized component ($n_{\rm H_2}=10^{6.5}$\,cm$^{-3}$) with log-uniform priors on its $T_{\rm kin}$ and $N_{\rm CO}/\Delta v$ and a smaller $\Omega$ does not improve the fit. Under these priors, at least three quarters of the CO(2\mbox{--}1) luminosity arises in the low-excitation gas (95\% probability), although warm gas could supply about half of the CO(5\mbox{--}4) luminosity.

\emph{Prior sensitivity.} Re-weighting the posterior to other priors changes the CMB contrasts, the corrected CO(1\mbox{--}0) luminosity, the \ci-based mass and the factor relative to thermalized gas by at most 16\%, except for near-virial gradients ($K_{\rm vir}\leq3$), which raise the CO(1\mbox{--}0) luminosity and the \ci-based mass by about a third; all medians stay within the adopted 16--84\% intervals (Table~\ref{tab:priors}).

\begin{table}[h]
\caption{Prior sensitivity of the LAE\mbox{-}2 non-LTE outputs.}\label{tab:priors}
\centering\footnotesize
\setlength{\tabcolsep}{2.2pt}
\renewcommand{\arraystretch}{1.15}
\begin{tabular}{lccccc}
\toprule
Prior & $C_{10}$ & $C_{\rm CI}$ & $\Lp_{\rm CO(1-0)}$\tablefootmark{a} & $\Mmol$\tablefootmark{b} & Factor\tablefootmark{c}\\
\midrule
Adopted & $0.63^{+0.20}_{-0.14}$ & $0.74^{+0.14}_{-0.22}$ & $6.7^{+3.6}_{-2.6}$ & $9.2^{+8.8}_{-4.3}$ & $4.0^{+2.5}_{-1.6}$\\
Flat in $T_{\rm kin}$ & $0.66^{+0.19}_{-0.13}$ & $0.77^{+0.12}_{-0.14}$ & $6.6^{+3.2}_{-2.3}$ & $8.5^{+7.1}_{-3.8}$ & $4.0^{+2.2}_{-1.4}$\\
$T_{\rm kin}\geq25$\,K & $0.65^{+0.20}_{-0.12}$ & $0.76^{+0.13}_{-0.19}$ & $6.4^{+3.2}_{-2.4}$ & $8.8^{+7.8}_{-4.1}$ & $3.9^{+2.2}_{-1.5}$\\
$n_{\rm H_2}\leq10^{4}$\,cm$^{-3}$ & $0.64^{+0.20}_{-0.14}$ & $0.76^{+0.13}_{-0.22}$ & $6.7^{+3.6}_{-2.6}$ & $9.1^{+8.6}_{-4.3}$ & $4.1^{+2.5}_{-1.6}$\\
$n_{\rm H_2}\geq10^{2.5}$\,cm$^{-3}$ & $0.70^{+0.19}_{-0.22}$ & $0.82^{+0.09}_{-0.28}$ & $5.6^{+3.6}_{-1.9}$ & $8.8^{+8.6}_{-4.1}$ & $3.4^{+2.4}_{-1.2}$\\
$K_{\rm vir}\leq3$ & $0.56^{+0.07}_{-0.10}$ & $0.55^{+0.12}_{-0.18}$ & $8.6^{+2.9}_{-2.4}$ & $12.3^{+11.6}_{-5.8}$ & $5.2^{+2.2}_{-1.6}$\\
Wider ${\rm d}v/{\rm d}r$\tablefootmark{d} & $0.63^{+0.21}_{-0.14}$ & $0.73$--$0.75$ & $6.6^{+3.6}_{-2.6}$ & $9.0$--$9.5$ & $4.0^{+2.5}_{-1.6}$\\
\bottomrule
\end{tabular}
\tablefoot{Medians and 16--84\% intervals after re-weighting; other priors unchanged. $C_{10}$ and $C_{\rm CI}$ are the CMB contrasts of CO(1\mbox{--}0) and \ci. $K_{\rm vir}$ for the density-profile coefficient $a=1$ of \citet{Papadopoulos2012} and $x_{\rm CO}=10^{-4}$. \tablefoottext{a}{CMB-corrected, $10^{10}$\,\Lpunit.} \tablefoottext{b}{\ci-based, $10^{10}\,\Msun$.} \tablefoottext{c}{CO(5\mbox{--}4) faintness per unit gas mass relative to thermalized 40\,K gas (see text).} \tablefoottext{d}{The full sampled range, ${\rm d}v/{\rm d}r=0.01$--$3\times10^{4}$\,\kms\,pc$^{-1}$, instead of 0.1--1000\,\kms\,pc$^{-1}$; the \ci\ entries bound the median.}}
\end{table}

\section{Dust and PDR modeling}\label{app:dustpdr}

\begin{figure*}[t]
\centering\includegraphics[width=180mm]{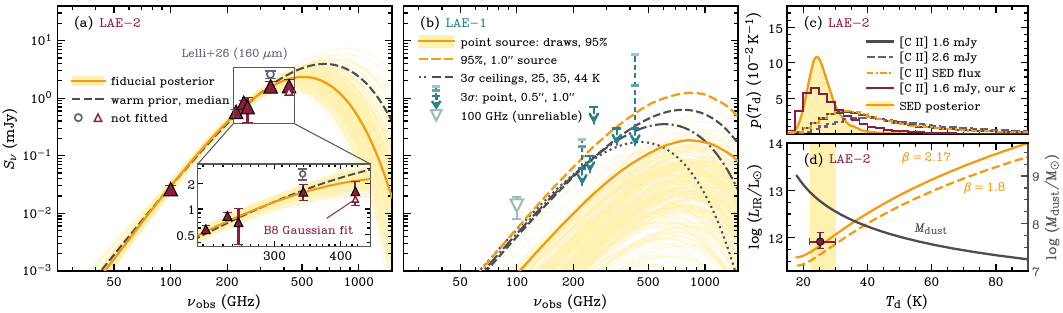}
\caption{Dust SEDs (a, b) and dust temperature of LAE\mbox{-}2 (c, d). (a) LAE\mbox{-}2: fitted fluxes (Table~\ref{tab:cont}) with the fiducial fit (200 draws) and the median for the ${\cal N}(44,6)$\,K prior; not fitted: the 426\,GHz Gaussian-fit flux and the 160\,$\mu$m flux of \citet{Lelli2026}. (b) LAE\mbox{-}1: limits for a point source and for 0.5\arcsec\ and 1.0\arcsec\ Gaussians (dark to light), the non-detection fit and modified blackbodies at the dust-mass limits; the 100\,GHz candidate sets none of them. (c) $T_{\rm d}$ from the SED fit and from the \cii\ estimator of \citet{Sommovigo2022}. (d) $\log L_{\rm IR}$ ($\beta=2.17$ and 1.8) and $\log M_{\rm dust}$ (right axis) vs. $T_{\rm d}$ for fixed-$T_{\rm d}$ fits, with the fiducial posterior (band and circle).}\label{fig:dust}
\end{figure*}

\begin{figure}[t]
\includegraphics[width=83mm]{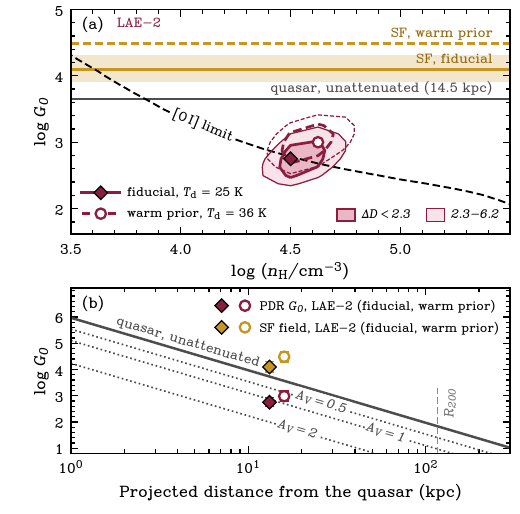}
\caption{PDR constraints for LAE\mbox{-}2 and the FUV fields at its position. (a) Deviance $\Delta D$ of the PDR fit above its minimum for the fiducial dust SED and the warm prior, with the best cells marked. Above the black dashed line, the model \oi/\cii\ ratio exceeds the \oi\ limit. The horizontal lines show the unattenuated quasar field at the projected 14.5\,kpc and the star-formation field $L_{\rm FIR}/(4A)$ of each dust fit (band: 16--84\% for the fiducial fit). (b) Quasar field vs. projected distance, unattenuated and attenuated by $e^{-2A_V}$, with the PDR and star-formation fields of LAE\mbox{-}2 (medians and 16--84\%) displaced from 14.5\,kpc for clarity. $R_{200}$ is the virial radius of a $10^{13}\,\Msun$ halo.}\label{fig:pdr}
\end{figure}

\emph{Continuum data.} At 223\,GHz (Table~\ref{tab:cont}), LAE\mbox{-}2 has a total flux of $579\pm34$\,$\mu$Jy, the same in a Gaussian fit and in an aperture of 0.8\arcsec\ radius, and a deconvolved FWHM size of $0.36\arcsec\times0.24\arcsec$; the quasar and the SMG give 5.4 and 5.9\,mJy, consistent with \citet{Lehnert2020}. At 426\,GHz, LAE\mbox{-}2 peaks at $8\sigma$ and is at most marginally resolved, with $1.62\pm0.45$\,mJy in a 0.5\arcsec\ radius aperture and 1.31\,mJy in a Gaussian fit, which we use as a sensitivity test. The source is far smaller than the maximum recoverable scale of either data set. The 223\,GHz flux agrees with the $830\pm10$\,$\mu$Jy of \citet{Lu2017} at 245\,GHz for $\beta\approx2$, so we find no evidence for substantial missed continuum flux; the 426/223\,GHz flux ratio also does not depend on the aperture once both images are smoothed to the same resolution. The $2.6\pm0.3$\,mJy of \citet{Lelli2026} at rest-frame 160\,$\mu$m, plotted at the mean frequency of their spectral windows (339.0\,GHz), is 60\% above the 340\,GHz flux but is summed within four continuum effective radii, an aperture that includes the tail toward the quasar.

\emph{Dust.} The fits constrain $L_{\rm IR}$ better than $T_{\rm d}$ and $\beta$, and none of those with free $\beta$ favors dust warmer than about 40\,K. We sampled the Gaussian likelihood with emcee, with a flat prior on the normalization, $\log(M_{\rm dust}/\Msun)\in[5,10.5]$. In the fiducial fit (Table~\ref{tab:scenarios}; Fig.~\ref{fig:dust}a), $\beta$ piles up toward its upper bound of 2.5, driven by the steep rise from 100 to 223\,GHz; without the 100\,GHz flux and with a flat $\beta$ prior of 1--3, $\beta=1.8^{+0.8}_{-0.6}$ and $T_{\rm d}\approx26$\,K. Values near the fiducial $\beta\approx2.4$ are measured in main-sequence galaxies at $z>5$ \citep{Faisst2020,Algera2024,Villanueva2024} and in laboratory analogs of cold amorphous silicates \citep{Demyk2017}. We repeated the fit with free $\beta$ for a finite opacity over the 4.28\,kpc$^2$ Gaussian area of the dust ($\tau_{158}=0.35^{+0.14}_{-0.12}$), for $\tau=1$ at 100--150\,$\mu$m, for $\beta$ up to 3, for the Gaussian-fit 426\,GHz flux, and for a 5\% Band~3 flux scale. These fits span $T_{\rm d}=18$--37\,K but keep $\log(L_{\rm IR}/\Lsun)$ within 11.84--11.94. Fixing $\beta$ at 2.17 or 1.8 gives 29 or 42\,K, with higher luminosities (12.02 and 12.39; Fig.~\ref{fig:dust}d) and poorer fits. The warm prior, ${\cal N}(44,6)$\,K, moves the posterior only to $T_{\rm d}=36^{+7}_{-6}$\,K, with a poorer fit (Table~\ref{tab:scenarios}), and fixing $T_{\rm d}=44$\,K raises $\chi^2$ from 2.4 to 12.3 (four instead of three degrees of freedom), because such dust would give a 426/223\,GHz flux ratio of 5.4--6.9, against the observed $2.8\pm0.9$. We do not fit the 160\,$\mu$m flux, which is in tension with the 223--257\,GHz fluxes; including it would raise $L_{\rm IR}$ by 0.1\,dex and shorten $\tdep$ from 0.72 to 0.57\,Gyr.

The \cii-based estimator of \citet{Sommovigo2022} gives $T_{\rm d}=41^{+25}_{-12}$\,K for the 340\,GHz flux (Fig.~\ref{fig:dust}c), but it assumes a Milky-Way dust opacity four times below ours at 158\,$\mu$m; with our $\kappa_\nu$, at a fixed GDR and \cii\ calibration, it gives $27^{+10}_{-6}$\,K, consistent with the SED fit.

\emph{Molecular gas mass from dust and metals.} The dust-continuum calibration of \citet{Dunne2022}, $\alpha_{850}=6.9\times10^{12}$\,W\,Hz$^{-1}$\,$\Msun^{-1}$, applied with the 25\,K mass-weighted temperature of \citet{Scoville2016} to the CMB-corrected 100\,GHz flux extrapolated with the fiducial $\beta$ (2.18--2.46), gives $\Mmol=(3.9$--$4.5)\times10^{10}\,\Msun$, lower than the \ci-based mass by a factor of 2.1--2.4. The \ci-based GDR is $380^{+450}_{-210}$ for the fiducial dust mass, whereas the near-solar ratio of \citet{RemyRuyer2014}, about 160 for their $\kappa_{850}$ and 246 for ours, would require $\log(M_{\rm dust}/\Msun)\gtrsim8.6$. If the nebular lines of LAE\mbox{-}2 are powered by star formation, its N2-based abundance, $12+\log({\rm O/H})=8.77$ \citep{Zamora2025}, and the closed-box oxygen yields of \citet{Vincenzo2016} for our IMF ($y_{\rm O}=0.027$--0.034) imply that the stars of LAE\mbox{-}2 can have enriched (7--10)$\times10^{10}\,\Msun$ of gas to this abundance, or (4--16)$\times10^{10}\,\Msun$ after propagating the abundance and stellar-mass uncertainties, consistent with the \ci-based mass. For $\aco=4.4$, the gas would nominally hold about three times the oxygen the stars can produce in a closed box. This check depends on the ionization source and calibration of the N2 abundance and is secondary to the \ci\ and \cii\ estimates.

\emph{LAE\mbox{-}1: dust limits.} After subtraction of Gaussian models of the quasar and the SMG, the 223\,GHz image gives total-flux limits at the extraction position of 60, 145 and 191\,$\mu$Jy for a point source and for 0.5\arcsec\ and 1.0\arcsec\ Gaussians (per-beam values in Table~\ref{tab:cont}); at the stellar peak LAE\mbox{-}1a, 0.5\arcsec\ away, the same image has $67\pm22$\,$\mu$Jy in a 0.5\arcsec\ aperture, below our $3.5\sigma$ detection threshold. We do not use the deeper Band~7 image of Fig.~\ref{fig:positions}, because no total flux at this position is published. The non-detection fit (Fig.~\ref{fig:dust}b) treats the forced 223 and 426\,GHz measurements at the extraction position as Gaussian amplitudes and the published non-detections as censored values at their limits; for extended sources, the 223\,GHz datum enters as the size-dependent limit. The fixed-$T_{\rm d}$ mass limits instead use the limit of each datum: for a point source with $\beta=1.8$, $\log(M_{\rm dust}/\Msun)<7.76$ and 7.14 at 25 and 44\,K, set by the 223 and 426\,GHz data, and for the two Gaussians the 245\,GHz limit of \citet{Lu2017} dominates and the mass limits rise by 0.24 and 0.48\,dex at 25\,K. Extrapolated to 223\,GHz, the 100\,GHz candidate (App.~\ref{app:spectra}) would be 2.7--3.5 times the point-source limit for 25--44\,K, a discrepancy of only $\approx2\sigma$. For dust on the 2--3\arcsec\ scale of this diffuse gas, beam dilution raises the mass limits by about 1\,dex, so the data do not constrain dust associated with it.

\emph{LAE\mbox{-}1: molecular gas.} With the GDRs of \citet{RemyRuyer2014} at the metallicity of LAE\mbox{-}1a and LAE\mbox{-}1b, 0.3--0.45\,$Z_\odot$ \citep{Zamora2025}, which are 551--796 for our $\kappa_{850}$, unresolved dust at 25\,K gives $\Mmol<(3.1$--$4.5)\times10^{10}\,\Msun$. For dust at LAE\mbox{-}1a or at the centroid of the pair, which our photometry recovers only partly, the 245\,GHz limit, assumed to apply across this region for uniform noise, gives $(4.2$--$6.0)\times10^{10}\,\Msun$. Because the excitation and the conversion factor of the gas at the extraction position are not known and it does not coincide with the stellar peaks (App.~\ref{app:spectra}), the limits on $\mu_{\rm gas}$ and $\tdep$ of LAE\mbox{-}1 (Table~\ref{tab:scenarios}) use the dust-based mass. The CO data do not reach the flux expected for LAE\mbox{-}1 on the adopted relation \citep{Daddi2015}: even ${\rm SFR_{IR}}=52\,\Msun$\,yr$^{-1}$ corresponds to a CO(5\mbox{--}4) flux of only 0.03\,\Jykms, half our $3\sigma$ sensitivity.

\begin{table}[t]
\caption{Ancillary lines of LAE\mbox{-}2 in the PDR and LVG fits.}\label{tab:anc}
\centering\footnotesize\setlength{\tabcolsep}{2.8pt}\renewcommand{\arraystretch}{1.0}
\begin{tabular}{lccccc}
\toprule
Line & Flux\un{\Jykms} & Width\un{\kms} & Aperture & Ref. & Fit\\
\midrule
\cii\,158\,$\mu$m & $5.3\pm0.5$ & total & $r=4R_{\rm e}$ & (1) & PDR\tablefootmark{a}\\
\nii\,205\,$\mu$m & $0.30\pm0.06$ & total ($\approx$1000) & -- & (2) & PDR\tablefootmark{a}\\
\oi\,145\,$\mu$m & $<0.79$ (1.33)\tablefootmark{b} & 338 (571) & 1.4\arcsec & (3) & PDR\\
CO(12\mbox{--}11) & $<0.32$ (0.54)\tablefootmark{b} & 338 (571) & 2.0\arcsec & (3) & LVG\\
\bottomrule
\end{tabular}
\tablefoot{(1) \citet{Lelli2026}; (2) \citet{Pavesi2016}; (3) \citet{Lee2021}. \tablefoottext{a}{The PDR fit uses the neutral \cii, $4.5\pm0.5$\,\Jykms, assuming that the total \nii\ flux, for which (2) states no aperture, covers the total \cii\ emission; their \cii/\nii\ ratio of 11 refers to the truncated \cii\ line of \citet{Carilli2013}.} \tablefoottext{b}{In parentheses: scaled linearly to the CO width.}}
\end{table}

\emph{PDR model.} We fitted the neutral \cii, \ci(1\mbox{--}0), CO(5\mbox{--}4), the \oi\,145\,$\mu$m limit (Table~\ref{tab:anc}) and $L_{\rm IR}$ with the Wolfire--Kaufman 2020 constant-density grids \citep{Wolfire2022} of the PDR Toolbox \citep{PoundWolfire2023} and a single effective area, $A_{\rm eff}$, with $L=4\pi A_{\rm eff}I$ for the tabulated face-on intensities $I$. The neutral \cii\ is the total line minus the ionized-gas contribution, $R_{\rm ion}L_{\rm [NII]}$ with $R_{\rm ion}=3.0$--4.3 \citep{Croxall2017}, or 84\% of the line for the total-line ratio $L_{\rm [CII]}/L_{\rm [NII]}=23\pm5$. Because the grids do not include the CMB, CO(5\mbox{--}4) and \ci\ are divided by the contrasts of 40\,K gas (0.80 and 0.77). The likelihood is Gaussian for the detected lines (with a 10\% absolute flux uncertainty), cumulative-Gaussian for the \oi\ limit, and uses the $L_{\rm IR}$ posterior of each dust fit (Table~\ref{tab:scenarios}); the prior is flat in $\log n_{\rm H}$ and $\log\Gnot$, and the area is optimized at each grid point. For the fiducial fit, $\log(n_{\rm H}/{\rm cm^{-3}})=4.54^{+0.10}_{-0.07}$ and $\log\Gnot=2.76^{+0.18}_{-0.17}$ (Fig.~\ref{fig:pdr}a; Table~\ref{tab:scenarios}), the luminosity reprocessed by the PDR component is $1.1^{+0.4}_{-0.3}$ times $L_{\rm IR}$, consistent with unity, and \oi\ is predicted at 0.8 times its limit. The fitted area, $205^{+29}_{-28}$\,kpc$^2$, is 69 times the FWHM area of the compact dust, so the PDR lines arise from the surfaces of many clouds or from gas more extended than the dust, as the extended \cii\ emission and the tail suggest \citep{Lelli2026}. For the warm prior, this ratio is $0.76^{+0.25}_{-0.19}$, and \oi\ is predicted at 1.1 times its limit. The cooling lines are therefore compatible with cold dust and a long $\tdep$. Scaled to the measured CO(5\mbox{--}4) with the 40\,K contrasts, the PDR component supplies only 31--37\% of the CO(2\mbox{--}1) luminosity (26\% with the fixed-gas factors of App.~\ref{app:radex}). The PDR component is thus a dense phase, about 30 times denser than the median one-component LVG solution (for $n_{\rm H}=2n_{\rm H_2}$), and less excited gas emits most of the CO(2\mbox{--}1).

\emph{Quasar field.} The unattenuated quasar field at LAE\mbox{-}2 exceeds the field required by the PDR lines, but a small fraction of the gas column would shield it. Pan-STARRS photometry of the quasar ($M_{1450}=-28.4$), integrated over 6--13.6\,eV with the slope of \citet{Lusso2015}, gives $L_{\rm FUV}=1.8\times10^{47}$\,erg\,s$^{-1}$, or $0.21\,L_{\rm bol}$ for $\log(L_{\rm bol}/{\rm erg\,s^{-1}})=47.93$ from the rest-frame ultraviolet continuum \citep{Yu2021,Zamora2025}; renormalizing this spectrum to the H$\alpha$- and 5100\,\AA-based values of 47.2 and 47.5 \citep{Zamora2025} would lower the field by factors of 5.4 and 2.6. In Habing units, the unattenuated field at 14.5\,kpc is $\Gnot=4500$ (Fig.~\ref{fig:pdr}b). Geometrically, this is an upper bound, because the true distance exceeds the projected one; for isotropic orientations, the median deprojection lowers $\Gnot$ by a factor of 1.3, and anisotropic emission from the accretion disk could change it by factors of 0.3--1.5. LAE\mbox{-}2 lies in projection inside the \lya\ halo, which \citet{Drake2020} tentatively link to the accreting black hole, but whether the quasar illuminates LAE\mbox{-}2 directly is not known. Attenuating this field to the fiducial PDR value would require $A_V=1.0\pm0.2$ for $\tau_{\rm FUV}=2A_V$, or $N_{\rm H}\approx1.9\times10^{21}$\,cm$^{-2}$ \citep{Bohlin1978}, only 0.6--2\% of the mean column of half the molecular gas within the radius adopted in App.~\ref{app:positions}, $(1.0$--$3.3)\times10^{23}$\,cm$^{-2}$. Shielded gas can still form stars, be redistributed tidally, or lose its exposed surfaces, so $\tdep$ measures consumption at the present SFR, not the lifetime of the clouds.

\emph{Star-formation field and dust heating.} We define the field of the galaxy's own star formation as $\Gnot^{\rm SF}=L_{\rm FIR}/(4A)$ in Habing units, the flux of half of the far-infrared luminosity (42.5--122.5\,$\mu$m) of each dust fit emerging through the two faces of the area $A$ (Table~\ref{tab:scenarios}). For $A$, we use the FWHM ellipse of the deconvolved 223\,GHz source, 2.97\,kpc$^2$ (circularized radius $R=0.97$\,kpc); the 2.0\,kpc radius from the size of \citet{Lu2017} would lower $\Gnot^{\rm SF}$ by 0.64\,dex. A disk of radius $R$ at the projected 14.5\,kpc intercepts $R^2/(4d^2)=0.11\%$ of the quasar output. Complete absorption of this fraction would supply at most $6\pm2\%$ (FUV) or $30^{+11}_{-10}\%$ (bolometric) of the fiducial $L_{\rm IR}$; for $R=2.0$\,kpc, these fractions are 4.3 times higher, so the bolometric bound is informative only for compact dust. Quasar heating is therefore not required to explain $L_{\rm IR}$, but it could supply up to about a third of the luminosity of the compact cold dust, in particular on the quasar-facing side.

\begin{table}[t]
\caption{Continuum data for LAE\mbox{-}1 and LAE\mbox{-}2.}\label{tab:cont}
\centering\footnotesize\setlength{\tabcolsep}{2.4pt}\renewcommand{\arraystretch}{0.95}
\begin{tabular}{lllll}
\toprule
$\nu_{\rm eff}$ & Data, beam & LAE\mbox{-}1 & LAE\mbox{-}2 & Scale\\
(GHz) & (\arcsec) & ($\mu$Jy) & ($\mu$Jy) & \\
\midrule
100 & B3\tablefootmark{a}, 1.6--2.0$\times$1.3--1.5 & $13.6\pm2.4\pm5.2$ & $26.4\pm2.3$ & 10\%\tablefootmark{g}\\
222.7 & B6\tablefootmark{b}, 0.55$\times$0.49 & $4\pm19$ & $579\pm34$ & 10\%\\
244.6 & B6\tablefootmark{c}, 1.0$\times$0.8 & $<100$ & $830\pm10$ & 10\%\\
256.6 & PdBI\tablefootmark{d}, 2.0$\times$0.9 & $<700$ & $700\pm300$ & 15\%\\
340 & B7\tablefootmark{e}, 1.2$\times$0.8 & $<300$ & $1600\pm350$ & 15\%\\
426.1 & B8\tablefootmark{f}, 0.27$\times$0.20 & $-58\pm98$ & $1620\pm450$ & 15\%\\
\bottomrule
\end{tabular}
\tablefoot{For LAE\mbox{-}1, the values at 100, 222.7 and 426.1\,GHz are our measurements at the extraction position, and the others are published limits. Errors are statistical except at 340\,GHz, where they include flux-scale uncertainty. Scale: flux-scale uncertainty, added in quadrature for LAE\mbox{-}2 (except at 340\,GHz) and marginalized for LAE\mbox{-}1.
\tablefoottext{a}{This work (App.~\ref{app:spectra}). LAE\mbox{-}1: statistical error and scatter among the four tunings; used only as an upper limit.}
\tablefoottext{b}{Our measurements on the data of \citet{Lehnert2020}. LAE\mbox{-}1: per beam (total-flux limits in the text); LAE\mbox{-}2: total flux (their published peak is $0.44\pm0.06$\,mJy\,beam$^{-1}$).}
\tablefoottext{c}{\citet{Lu2017}. LAE\mbox{-}1: point-source limit, position not tabulated; LAE\mbox{-}2: Gaussian fit.}
\tablefoottext{d}{\citet{Decarli2014}.}
\tablefoottext{e}{LAE\mbox{-}1: \citet{Carilli2013}; LAE\mbox{-}2: mean of \citet{Wagg2012} and \citet{Carilli2013}, with an error that spans both.}
\tablefoottext{f}{Our measurements on the data of \citet{Lee2019}. LAE\mbox{-}1: per beam; LAE\mbox{-}2: aperture of 0.5\arcsec\ radius.}
\tablefoottext{g}{Twice the nominal 5\%.}}
\end{table}

\begin{table}[t]
\caption{Derived properties of LAE\mbox{-}2 and LAE\mbox{-}1.}\label{tab:scenarios}
\centering\footnotesize\setlength{\tabcolsep}{3pt}\renewcommand{\arraystretch}{1.05}
\begin{tabular}{lccc}
\toprule
 & Fiducial & Finite opacity & Warm prior\\
\midrule
$T_{\rm d}$\un{K}\tablefootmark{a} & $25^{+5}_{-3}$ & $29^{+4}_{-3}$ & $36^{+7}_{-6}$\\
$\beta$ & $2.36^{+0.10}_{-0.19}$ & $2.37^{+0.10}_{-0.18}$ & $2.06^{+0.21}_{-0.19}$\\
$\chi^2$ / dof\tablefootmark{b} & 2.4 / 3 & 1.8 / 3 & 8.6 / 3\\
$\log L_{\rm IR}$\un{$\Lsun$} & $11.92^{+0.18}_{-0.13}$ & $11.84^{+0.18}_{-0.14}$ & $12.27\pm0.20$\\
$\delta_{\rm MS}$\un{dex} & ${+}0.12^{+0.18}_{-0.13}$ & ${+}0.05^{+0.18}_{-0.14}$ & ${+}0.47\pm0.20$\\
$\tdep$\un{Gyr} & $0.72^{+0.80}_{-0.38}$ & $0.87^{+0.97}_{-0.46}$ & $0.34^{+0.41}_{-0.18}$\\
$\Delta_{54}$\un{dex}\tablefootmark{c} & ${+}0.59^{+0.14}_{-0.19}$ & ${+}0.67^{+0.15}_{-0.19}$ & ${+}0.24^{+0.20}_{-0.21}$\\
$\log\Gnot^{\rm SF}$\tablefootmark{d} & $4.10^{+0.22}_{-0.19}$ & $4.00^{+0.23}_{-0.20}$ & $4.49^{+0.18}_{-0.20}$\\
$\log\Gnot^{\rm PDR}$\tablefootmark{d} & $2.76^{+0.18}_{-0.17}$ & $2.71^{+0.20}_{-0.18}$ & $2.99^{+0.18}_{-0.17}$\\
$\log M_{\rm dust}$\un{$\Msun$} & $8.38^{+0.19}_{-0.17}$ & $8.24\pm0.12$ & $8.11^{+0.13}_{-0.12}$\\
\midrule
\multicolumn{4}{l}{\emph{LAE\mbox{-}1}\tablefootmark{e}}\\
$\log L_{\rm IR}$\un{$\Lsun$} & \multicolumn{3}{c}{$<11.55$}\\
${\rm SFR_{IR}}$\un{$\Msun$\,yr$^{-1}$} & \multicolumn{3}{c}{$<52$}\\
$\Mmol$\un{$10^{10}\,\Msun$} & \multicolumn{3}{c}{$<3.1$--4.5 (dust); 0.55--2.2 (CO)\tablefootmark{f}}\\
$\mu_{\rm gas}$ & \multicolumn{3}{c}{$<55$}\\
$\tdep$\un{Gyr} & \multicolumn{3}{c}{$<5.0$}\\
\bottomrule
\end{tabular}
\tablefoot{Fits to the six fluxes of Table~\ref{tab:cont}, including CMB heating and contrast \citep{daCunha2013}. Medians and 16--84\% intervals, with $\Mstar$ fixed for $\delta_{\rm MS}$; $\tdep$ uses the \ci-based gas mass.
\tablefoottext{a}{Intrinsic (without CMB heating).}
\tablefoottext{b}{At the posterior mode.}
\tablefoottext{c}{Offset from the $\Lp_{\rm CO(5-4)}$--$L_{\rm IR}$ relation of \citet{Daddi2015} for the observed CO(5\mbox{--}4).}
\tablefoottext{d}{In Habing units; $\Gnot^{\rm PDR}$ is the marginal posterior on the 0.125\,dex grid.}
\tablefoottext{e}{Limits for a point source at the extraction position, with 44\,K for $L_{\rm IR}$ and ${\rm SFR_{IR}}$ and 25\,K for the dust-based $\Mmol$; $\mu_{\rm gas}$ and $\tdep$ are for LAE\mbox{-}1 as a whole and use this mass with $\log(\Mstar/\Msun)=8.92$ and ${\rm SFR}=9\,\Msun$\,yr$^{-1}$ (App.~\ref{app:comparison}). For dust at the stellar positions, $\Mmol<6.0$, $\mu_{\rm gas}<73$ and $\tdep<6.7$.} \tablefoottext{f}{From the CO(5\mbox{--}4) line at the extraction position for $\aco=0.8$, between thermalized 40\,K gas and the excitation of LAE\mbox{-}2.}}
\end{table}

\section{Gas fractions and depletion times}\label{app:comparison}

\begin{figure}[t]
\centering\includegraphics[width=82mm]{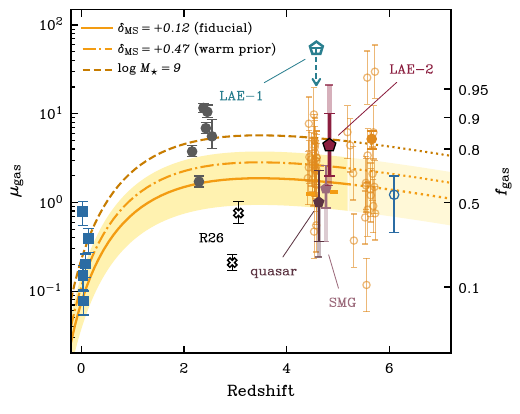}
\caption{Gas-to-stellar mass ratio $\mu_{\rm gas}$ vs. redshift (right axis: $f_{\rm gas}=\mu_{\rm gas}/(1+\mu_{\rm gas})$). Curves, bars, and comparison symbols follow Fig.~\ref{fig:consumption}b. Pentagons: LAE\mbox{-}2, quasar host and SMG, using \ci\ gas masses and the SED stellar masses of \citet{Lelli2026}. Orange ticks: the relation at the quasar/SMG stellar masses and IR-based $\delta_{\rm MS}$. LAE\mbox{-}1: limit from the 25\,K dust-based gas mass and $\log(\Mstar/\Msun)=8.92$. Crosses: medians inside (lower) and outside (upper) quasar virial radii at $z\approx3$ \citep[App.~\ref{app:comparison};][]{Ritter2026}.}\label{fig:gasfraction}
\end{figure}

\emph{Comparison samples.} Comparison masses in Figs.~\ref{fig:consumption} and \ref{fig:gasfraction} use different tracers and conversion factors. CO-based values adopt the published $\aco$: metallicity-dependent (4.5--11.7) for LARS \citep{Puschnig2020}, 4.5 for HZ10 at $z=5.65$ \citep{Pavesi2019} and J0235$-$0532 \citep{Ono2022}, 1 for the UV-bright \lya\ emitters at $z\approx2.5$ \citep{Dessauges2025} and for CRLE at $z=5.67$, on a modeled CO(1\mbox{--}0) \citep{Vieira2022}, and 0.8 with $r_{31}=1$ for LABd05 \citep{Yang2014}. The ALPINE masses (open circles) come from \cii\ using the calibration of \citet{Zanella2018} \citep{Bethermin2020,Dessauges2020}, and those of the $z>6$ quasar companions come from dust, with a GDR of 100 \citep{Pensabene2021}. Offsets of a factor of 3--5 between these methods are therefore expected.

\emph{Conventions of the comparison.} Error bars are the published ones, combined in log space for $\tdep$, or 0.3\,dex where none are tabulated (LARS, ALPINE). We retain the published IMFs and cosmologies of the comparison samples; where stated, each differs from ours by $\lesssim8$\% in mass and SFR. The samples are effectively CMB-corrected or insensitive to the CMB, except HZ10, whose $\aco$ includes the CMB, and CRLE and LABd05, whose CMB treatment is not stated. In Fig.~\ref{fig:consumption}, the group members are shifted by $|\Delta z|\leq0.12$ for clarity, J0235$-$0532 is a $4\sigma$ detection, and the open diamonds are the $z=1.1$--1.6 sample of \citet{Valentino2020}.

\emph{Stellar masses and SFRs.} The AGN-inclusive SED fits of \citet{Lelli2026} give $\log(\Mstar/\Msun)=10.33^{+0.20}_{-0.22}$, $11.46^{+0.45}_{-0.35}$ and $11.45^{+0.21}_{-0.26}$ for LAE\mbox{-}2, the quasar host and the SMG. We adopt these masses and the SED-based SFRs as published; their IMF is not stated. If it is Salpeter, conversion to our IMF multiplies both by about 0.66, and $\mu_{\rm gas}$ and the SED-based $\tdep$ by 1.5 \citep{Madau2014}. For LAE\mbox{-}1, the limits use $\log(\Mstar/\Msun)=8.92$, the lower end of the 8.9--9.4 range of \citet{Fontana1996} rescaled to our IMF and luminosity distance; our fit to the HST and JWST photometry of the pair gives $9.34^{+0.14}_{-0.08}$, within this range. The UV/\lya\ SFR of LAE\mbox{-}1, 9--23\,$\Msun$\,yr$^{-1}$ in our IMF and cosmology \citep{Ohyama2004,Williams2014,Drake2020}, may include halo light.

\emph{Quasar host and SMG.} The quasar host and the SMG hold more molecular gas than LAE\mbox{-}2 and have shorter median $\tdep$ on either SFR basis. We derive their gas masses from our \ci\ fluxes with the same $\alpha_{\rm CI}$ and $T_{\rm ex}=40$\,K (error bars for 30--60\,K), which gives 3--4 and 4--5 times the gas mass of LAE\mbox{-}2; the pale bars in Fig.~\ref{fig:consumption}b extend down to the masses for $\aco=0.8$ applied to the CMB-corrected CO(2\mbox{--}1) luminosities of \citet{Wagg2014}. If star formation heats all their dust, the 20--1000\,$\mu$m luminosities of \citet{Lu2017} give SFRs of 4300 and 4500\,$\Msun$\,yr$^{-1}$ and $\tdep$ of 0.07 and 0.09\,Gyr, 8--11 times shorter than in LAE\mbox{-}2. The SED-based SFRs of \citet{Lelli2026}, $1000^{+1300}_{-750}$ and $740^{+1260}_{-280}\,\Msun$\,yr$^{-1}$, which average over the last 100\,Myr, give $0.30^{+0.89}_{-0.17}$ and $0.52^{+0.36}_{-0.32}$\,Gyr, against $0.88^{+1.5}_{-0.6}$\,Gyr for LAE\mbox{-}2 on the same basis. Relative to the $\Lp_{\rm CO(5-4)}$--$L_{\rm IR}$ relation of \citet{Daddi2015}, the two galaxies lie at $-0.22$ and $-0.06$\,dex on the IR basis and at $+0.4^{+0.6}_{-0.3}$ and $+0.7^{+0.2}_{-0.4}$\,dex on the SED basis (open pentagons in Fig.~\ref{fig:consumption}a).

\emph{Gas-to-stellar mass ratios.} With the \ci-based gas masses, the quasar host and the SMG have $\mu_{\rm gas}=1.0^{+1.3}_{-0.6}$ and $1.4^{+1.2}_{-0.5}$ (Fig.~\ref{fig:gasfraction}); the $f_{\rm gas}$ of LAE\mbox{-}2 spans 0.60--0.95 over the $\aco$ range of Sect.~\ref{sec:gas}. Relative to the relation of \citet{Tacconi2020} at their stellar masses and main-sequence offsets, LAE\mbox{-}2, the quasar host, and the SMG lie at $+0.41$, $-0.10$, and $+0.04$\,dex for IR-based offsets. \citet{Ritter2026} use $\Mstar=0.88\times3.1\times10^{5}\,{\rm FWHM}^{2}R-\Mmol$, a CO-linewidth dynamical estimate minus the gas mass (average inclination, 12\% dark matter), with $\aco=0.8$ applied here to our modeled, CMB-corrected CO(1\mbox{--}0); their masses use observed CO(4\mbox{--}3) with adopted line ratios. For LAE\mbox{-}2, $\Mmol/\Mstar=0.20$--0.43 is comparable to or above their median of 0.21 inside $z\approx3$ quasar virial radii, but below 0.76 outside.

\end{appendix}
\end{CJK*}
\end{document}